\documentclass[aps,prb,twocolumn,showpacs,amsmath,amssymb,superscriptaddress,floatfix]{revtex4-1}
\usepackage{epstopdf}
\usepackage{graphicx}
\usepackage{dcolumn}
\usepackage{bm}
\usepackage{color}
\begin{document}

\title{Ruderman-Kittel-Kasuya-Yosida interaction in biased bilayer graphene}
\date{\today}

\author{F. Parhizgar}
\affiliation{School of Physics, Institute for Research in
Fundamental Sciences (IPM), Tehran 19395-5531, Iran}
\author{M. Sherafati}
\affiliation{Department of Physics $\&$ Astronomy, University of Missouri,
Columbia, MO 65211, USA}
\author{R. Asgari}
\email{asgari@ipm.ir} \affiliation{School of Physics, Institute
for Research in Fundamental Sciences (IPM), Tehran 19395-5531,
Iran}
\author{S. Satpathy}
\affiliation{Department of Physics $\&$ Astronomy, University of Missouri,
Columbia, MO 65211, USA}


\begin{abstract}

We study the Ruderman-Kittel-Kasuya-Yosida (RKKY) interaction between two contact magnetic impurities placed on bilayer graphene (BLG). We compute the interaction mediated by the carriers of the pristine and biased BLG as well as the conduction electrons of the doped system. The results are obtained from the linear-response expression for the susceptibility written in terms of the integral over lattice Green's functions. For the unbiased system, we obtain some analytical expressions in terms of the Meijer G-functions, which consist of the product of two oscillatory terms, one coming from the interference between the two Dirac points and the second coming from the Fermi momentum. In particular, for the undoped BLG, the system exhibits the RKKY interaction commensurate with its bipartite nature as expected from the particle-hole symmetry of the system. Furthermore, we explore a beating pattern of oscillations of the RKKY interaction in a highly doped BLG system within the four-band continuum model. Besides, we
discuss the discrepancy between the short-range RKKY interaction calculated from the two-band model and that obtained from the four-band continuum model. The final results for the applied gate voltage are obtained numerically and are fitted with the functional forms based on the results for the unbiased case. In this case, we show that the long-range behavior is scaled with a momentum that depends on Fermi energy and gate voltage, allowing the possibility of tuning of the RKKY interaction by gate voltage.
\end{abstract}
\maketitle

\section{Introduction}

Graphene, a two-dimensional (2D) honeycomb lattice of carbon atoms, was thrust into the limelight of condensed-matter research since its experimental emergence in 2004 \cite{Novoselov,Castro Neto}. Much of this attraction is due to its 2D structure and contrary to any ordinary 2D material, having two Dirac cones in the Brillouin zone (BZ), where the conduction and valence bands touch. Charge carriers with momenta near these two cones (known as $\bm K$ and $\bm K'$) have a unique linear energy dispersion and behave like massless Dirac fermions. On the other hand, crystalline BLG \cite{berger,novo,zhang} has recently attracted a great deal of attention because of its unique tunable electronic properties. It consists of two single-layer graphene (SLG) sheets separated by a small distance and can be produced by mechanical exfoliation of thin graphite or by thermal decomposition of silicon carbide. The low-energy quasi-particles in BLG behave as massive chiral fermions and are responsible for a plethora of
interesting physics including broken-symmetry states at very weak magnetic fields when BLG is suspended to reduce disorder~\cite{vafek} and anomalous exciton condensation in the quantum Hall regime~\cite{bilayer2}. Although the intrinsic BLG is a zero-gap semi-metal, it becomes a tunable band gap semiconductor~\cite{mak, kuzmenko} when a gate voltage is applied. The band gap determines the threshold voltage and the on-off ratio of field-effect transistors and diodes, thereby making BLG more convenient for applications in nano-electronic industry than SLG ~\cite{bilayer1, Nanda-Bilayer}.

One of the fundamental problems of interest in graphene research is the indirect exchange interaction between two localized magnetic moments placed on this otherwise non-magnetic material. This carrier-mediated exchange interaction is known as RKKY interaction \cite{Ruderman,Kasuya,Yosida} and it plays a significant role in the magnetic ordering of many electronic systems including spin glasses and alloys. As it was originally studied for three-dimensional electron gas, it has also been studied for electron gas in one \cite{RKKY-1D} and two \cite{RKKY-2D} dimensions. Two main features of the long-range behavior of the interaction, measured by exchange integral, $J$, for an electron gas is that it oscillates (in sign and magnitude) with the distance, $R$, between the moments, which exhibits ferromagnetic (FM) or anti-ferromagnetic (AFM) ordering and also decays~\cite{RKKY-1D,RKKY-2D} with $R$. Both of these features have different functional forms depending on the dimension and generally, on the energy
dispersion of the host material. For SLG, the RKKY interaction has extensively been studied \cite{Vozmediano,Dugaev,Saremi,Brey,Bunder,Black-Schaffer,fariborz,MSashi1,MSashi2,Kogan,MSashi3}. For an undoped SLG ($E_F=0$) two main features are agreed upon, first, unlike an ordinary 2D metal with $R^{-2}$ decay in the long-distance limit, $J$ in undoped graphene falls off as $R^{-3}$ and shows the $1 + \cos [(\bm{K}-\bm{K'})\cdot \bm{R}]$-type oscillations with additional phase factors~\cite{MSashi1} depending on the direction of $\bm{R}$, and second, the moments on the same sublattice exhibit an FM interaction and an AFM coupling if placed on the opposite sublattices, as required by the particle-hole symmetry~\cite{Saremi}. The RKKY interaction for doped graphene shows a long-range behavior similar to that of ordinary 2D electron gas with another oscillatory factor emerging from the Dirac cones. It was shown that two characteristic momenta, $\bm{k}_F$ and $\bm{K}-\bm{K'}$ can be tuned to exhibit an unusual
beating of the RKKY interaction for certain magnetic moment arrangements \cite{MSashi2}.

The RKKY interaction in BLG has also been addressed by several researchers~\cite{Hwang, Killi, Kogan, Jiang}. The local moment formation for adatoms on BLG using a mean-field theory of the Anderson impurity model has been studied by Killi et al~\cite{Killi}. They showed that the RKKY interaction between local moments can be varied by tuning the chemical potential or by tuning the electric field as it induces changes in the band structure of BLG. The symmetry of the RKKY interaction on the bipartite lattice at half filling has been discussed recently~\cite{Kogan} and the distance dependence of the RKKY interaction has been briefly reported. Furthermore, Jiang et al~\cite{Jiang}, investigated the RKKY interaction in multilayer graphene systems and they showed that the thickness of the multilayer influences the interaction in a complicated manner and that the interaction couplings fall off as $R^{-2}$ in long-range regime for BLG.

However, the previous studies have only considered the RKKY interaction in the half-filled ($E_F=0$) BLG. Consequently, significant tunability feature of the RKKY interaction due to both doping, where the Fermi energy is no longer zero, and the perpendicular electric field, which gives rise to the gapped BLG, have not been addressed in the literature. Both of these cases are of paramount importance when it comes to possible spintronic applications and will be the main focus in the present work.

In this paper, we extend the Green's function (GF) technique used for the RKKY interaction in SLG~\cite{MSashi1,MSashi2} to BLG. All cases of undoped, doped, unbiased and biased system are considered. We use the effective two-band Hamiltonian for the BLG~\cite{McCann, Nilsson, Castro} and for the first time, report the analytical expressions of the RKKY interaction for unbiased BLG in terms of the Meijer G-functions. We also present the numerical results of the interaction in the presence of a perpendicular electric field and show how the long-range behavior of the interaction can be tuned by the gate voltage. We explore a beating pattern of oscillations of the RKKY interaction in the four-band continuum model in which two conduction bands are partially occupied ( highly doped system). Furthermore, we discuss the discrepancy between the short-range RKKY interaction calculated from the two-band and that obtained from the four-band continuum model.

The paper is organized as follows. In Sec. II, we introduce the general formalism and all the required GFs that will be used in calculating the RKKY interaction within the GF approach. In Sec. III and IV, we present the analytical results of the RKKY coupling within the two-band and four-band models, respectively. The main numerical results using two-band and four-band models and a brief account of the difference between the RKKY interaction within the four-band model and the effective two-band approximation are presented in Sec. V. Finally, we summarize the results in Sec. VI and draw some conclusions. We have used the same GF method in studying the impurity states induced by a single vacancy in SLG, that includes the behavior of the $\sigma$ electrons as well as the $\pi$ electrons both using model and density-functional calculations.~\cite{Nanda, MSashi5}

\section{The Model Hamiltonian and Formalism}

\begin{figure}
\includegraphics[width=1.0\linewidth]{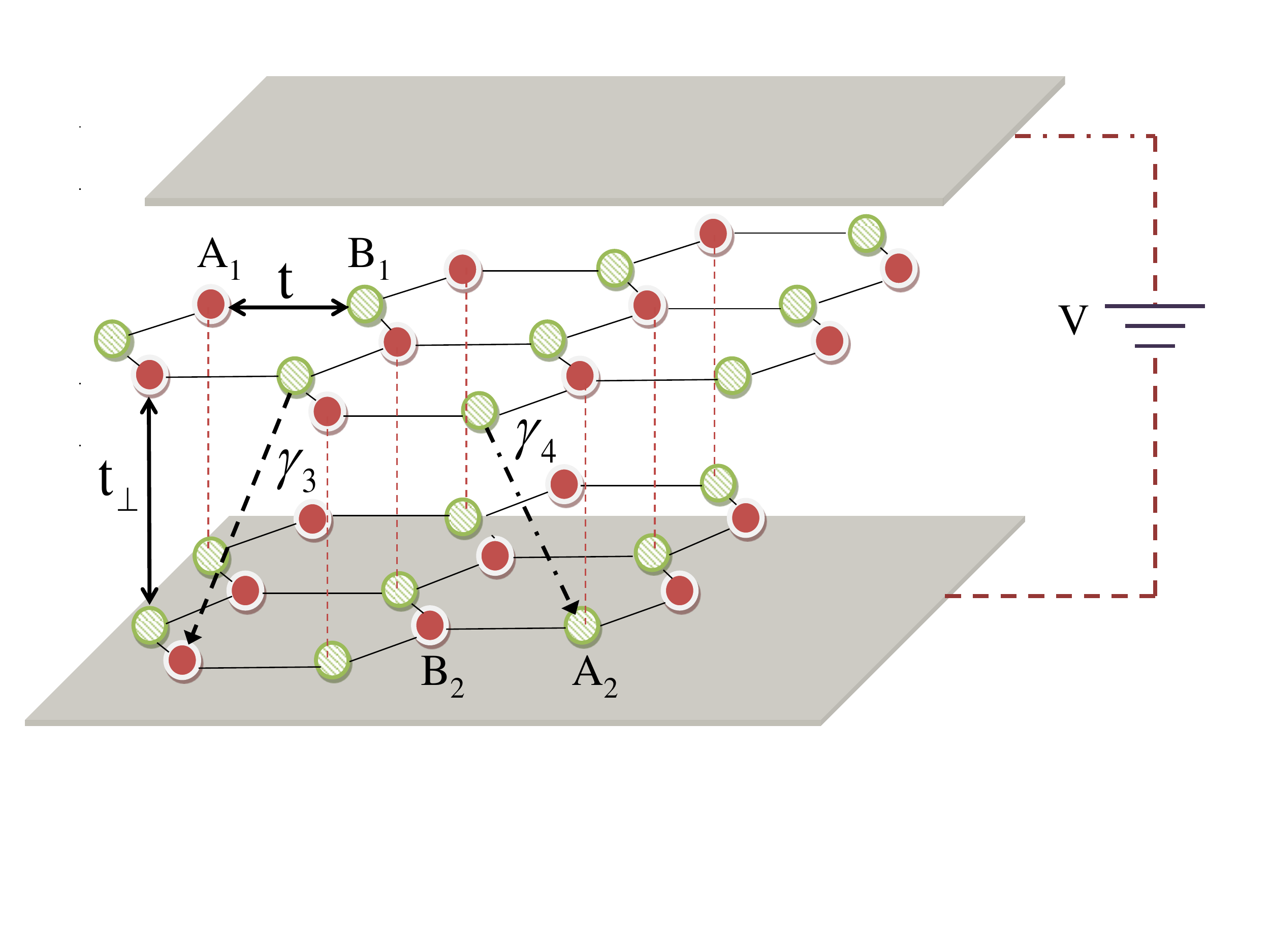}
\caption{ (Color online) Lattice structure of the BLG. The $A$ ($B$) sublattices are indicated by red (green) circles with corresponding intra-layer and inter-layer hopping amplitudes. The bias voltage is denoted by $V$.}
\label{fig:BLG}
\end{figure}

The BLG in Bernal stacking lattice shown in Fig.~\ref{fig:BLG} consists of two SLG lattices offset from each other in the $xy$ plane with four atoms in the unit cell such that the top $A$-sublattice is directly above the bottom $A$-sublattice and it is between these pairs of atoms that the inter-layer dimer bonds are formed. The other two atoms do not have a counterpart on the other layer. We assume that the $sp^2$-hybridized electrons of carbon atoms in each sheet are inert and only take into account the $2p_z$ electrons which form the $\pi$ bands.

We consider two magnetic impurities located at $(\alpha,0)$ and $(\beta,\bm{R})$ and in contact interaction with the electrons of the biased BLG in Bernal stacking, where $\alpha$ and $\beta$ denote the sublattice indices ($=A_1, B_1, A_2, B_2$). The tight-binding Hamiltonian of the system is given by
\begin{eqnarray}
{\cal H}  ={\cal H}_0+{\cal H}_{\text{int}},
\label{eq:hamil}
\end{eqnarray}
where the Hamiltonian for the biased BLG, ${\cal H}_0$, is given by

\begin{eqnarray}
{\cal H}_0=&&\frac{V}{2}\sum_{l=1,2}(-1)^{l+1}\sum_{i,\alpha=A,B} c^\dagger _{\alpha_l,\bm R_i}c_{\alpha_l,\bm R_i}  \nonumber\\
&&-t\sum_i  \sum_{j=1-3\,,l=1,2} c^\dagger _{A_l,\bm R_i}  c_{B_l,\bm R_i+\bm{\delta} _j} \nonumber \\
&&-t_\perp \sum_{i} c^\dagger _{A_1 ,\bm R_i} c_{A_2,\bm R_i}- \gamma_3 \sum_{i,j} c^\dagger _{B_1 ,\bm R_i} c_{B_2,\bm R_i+\delta_j}\nonumber\\
&& -\gamma_4 \sum_{i,j}( c^\dagger _{A_2 ,\bm R_i} c_{B_1,\bm R_i+\bm{\delta}_j}+ c^\dagger _{A_1 ,R_i} c_{B_2,\bm R_i+\bm{\delta}_j})+H.C,
\label{eq:H0TB}
\end{eqnarray}
where $l$ is the layer index, $V$ is an external potential difference between the layers, $t = 2.9$ eV is the intra-layer nearest-neighbor hopping energy \cite{Malard}, the hopping energy between on-top sublattices in different layers is $t_\perp = 0.3$ eV and furthermore, $\gamma_{3} = 0.12 \ eV$ denotes the hopping energy between not on-top sublattices between two layers~\cite{kuzmenko}. Another inter-layer second-nearest-neighbor hopping energies $\gamma_{4} = 0.04$ eV and hence very small compared to $t$ and can be ignored. The position vectors of three nearest neighbors of $A$-atom is denoted by $\bm{\delta}_j$ and $a \sim 1.4 \ {\AA}$ is the carbon-carbon bond length. In the most general case, the on-site energies on the four atomic sites are no longer equal. They consist of independent parameters to describe inter-layer asymmetry between the layers, an energy difference between two atoms in each layer and finally an energy difference between dimer and non-dimer sites. However, in this paper, we assume
the equal on-site energies. The wave function can be written as a four-component spinor, ${\psi_{A_1}, \psi_{B_1}, \psi_{A_2}, \psi_{B_2}}$. In this basis, the Hamiltonian of the biased BLG in Eq. \eqref{eq:H0TB} is represented as a $4\times4$ matrix given by ~\cite{Nilsson}
\begin{eqnarray}\label{4band}
{\cal H}_0=\begin{pmatrix} V/2 & f(\bm k) & t_\perp & v_4 f^*(\bm k)\\
f^*(\bm k) & V/2 & v_4f^*(\bm k) & v_3f(\bm k)\\
t_\perp & v_4f(\bm k) & -V/2 & f^*(\bm k)\\
v_4f(\bm k) & v_3f^*(\bm k) & f(\bm k) & -V/2 \end{pmatrix}
\label{eq:H04bnd}
\end{eqnarray}
where $f(\bm k)=-t\sum_{i=1}^3 e^{i{\bm k \cdot \bm{\delta}_i}}$, $v_3=\gamma_3/t$ and $v_4=\gamma_4/t$. The interaction between the localized spins $\bm{S}_1$ and $\bm {S}_2$ and the itinerant electron spins $\bm{s}$ is given by
\begin{eqnarray}
{\cal H}_{\text{int}} = - \lambda (\bm{S}_1 \cdot \bm{s}_1+\bm{S}_2 \cdot \bm{s}_2).
\end{eqnarray}

Ignoring $\gamma_3$ and $\gamma_4$ in the Hamiltonian justifies the use of the BZ for SLG to describe the electrons momenta for the BLG. Therefore, we similarly describe the physics for those electrons with momenta in the proximity of the Dirac points $\bm K_D=\bm K, \bm K'$. In order to find the low-energy Hamiltonian near Dirac points, we expand the function $f(\bm k)$ using $\bm k=\bm q+\bm K_D$ in powers of $\bm q$ and keep only the linear term, which yields $f(\bm k)=f(\bm q+\bm K_D) \simeq v_F q e^{i s\theta_q}$ where $s=\pm$ indicates the valley label, $v_F=3ta/2$ ( $\hbar=1$ from now on) is the Fermi velocity of the electrons in SLG, $q=\sqrt{q^2_x+q^2_y}$ and $\theta_q=\tan^{-1}(q_y/q_x)$ is the polar angle of $\bm q$ with respect to the $x$-axis chosen to be along the direction of $\bm K-\bm K'$. Furthermore, we consider $v_Fq,V\ll t_\perp$ which allows us to eliminate the high-energy states perturbatively and simplify the Hamiltonian in Eq. \eqref{eq:H04bnd} to two-band effective Hamiltonian with states localized around
$B_1$ and $B_2$ sites. Doing so yields four bands, two degenerate bands in each of the two valleys $\bm K$ and $\bm K'$, described by the effective two-band Hamiltonian~\cite{McCann}
\begin{eqnarray}
H_0=\frac{-1}{2m}\,\begin{pmatrix}0&q^2e^{-2i s\theta_q}\\q^2e^{2i s\theta_q}&0\end{pmatrix} +\begin{pmatrix}V/2&0\\0&-V/2\end{pmatrix},
\label{eq:H02bnd}
\end{eqnarray}
where $s=+1$ for the $\bm K$ valley and $s=-1$ for $\bm K'$ valley and $m=t_\perp/(2v_F^2)$ and it is about $0.03 m_e$ corresponding to a very small effective mass. The spinor is defined as $\psi^{\dagger}=(a^{\dagger}_{B_1},a^{\dagger}_{B_2})$ where $a^{\dagger}_{B_1}$ creates an electron mostly at the $B_1$ site with a small admixture from the other sites. We emphasize that the unperturbed Hamiltonians for the four-band and two-band models are denoted by ${\cal H}_0$ and $H_0$, respectively.

\subsection{The RKKY interaction $J_{\alpha,\beta}(\bm R)$}

In the linear-response theory, the strength of the RKKY interaction, $J$, is found by two steps. First, using the Lippmann-Schwinger equation $\vert\Psi\rangle=\vert\Psi^0\rangle+G^0 V\vert\Psi\rangle$, one calculates the perturbed state $\vert\Psi\rangle$, of the surrounding electron gas (host material) at the unperturbed state $\vert\Psi^0\rangle$ due to the first moment, $\bm{S}_1$ localized at the origin and second, the first-order correction in the energy of this spin-polarized gas is found in the presence of the second moment, $\bm{S}_2$ localized at the lattice position $\bm{R}$, viz., $E(\bm{R})=\langle\Psi\vert V(\bm{R}) \vert\Psi\rangle$. Here, $G^0=(E+i 0^{+}-{\cal H}_0)^{-1}$ is the the unperturbed retarded GF. Therefore, the interaction energy may be written as
\begin{eqnarray}
E(\bm{R}) = J(0,\bm{R}) \bm {S}_1 \cdot \bm {S}_2,
\label{eq:RKKYE}
\end{eqnarray}
with the RKKY interaction $J(0,\bm{R})$ being proportional to the static susceptibility, $\chi(0,\bm{R})$ viz.,
\begin{eqnarray}
J(0,\bm{R}) = \frac{\lambda^2 }{4} \chi (0,\bm{R}),
\label{eq:RKKYJ}
\end{eqnarray}
where the static susceptibility measures the proportionality between the perturbation $\delta V$ and the resulting change in the density $\delta n$, viz., $\chi(\bm{r},\bm{r'})=\delta n(\bm{r})/\delta V(\bm{r'})$.

It can be shown that $\chi(\bm{r},\bm{r'})$ is written as~\cite{Mohn}
\begin{eqnarray}
\chi(\bm{r},\bm{r'}) =
- \frac{2}{\pi} \int^{E_{\rm F}}_{-\infty} dE \
\Im m[G^0(\bm{r}, \bm{r'}, E)   G^0 (\bm{r'},\bm{r}, E)],
\label{eq:chiGG}
\end{eqnarray}
where $G^0(\bm{r}, \bm{r'}, E)=\sum_\mu \psi_\mu (\bm{r})\psi_\mu^*(\bm{r'})(E+i0^{+}-E_\mu)^{-1}$ is the real-space matrix element of the retarded GF for a single spin channel with $\mu$ labelling the complete set of eigenstates of ${\cal H}_0$. The factor $2$ behind the integral counts for both spin channels. Eq. \eqref{eq:chiGG} is obtained by using the relationship between the charge density and the perturbed GF, viz., $n(\bm{r})= \sum^{\text{occ}}_\mu\vert\psi_\mu (\bm{r})\vert^2=- \frac{2}{\pi} \int^{E_{\rm F}}_{-\infty} dE\ \Im m \ G(\bm{r},\bm{r}, E)$ and obtaining the charge difference $ \delta n(\bm{r})= n(\bm{r})- n^0(\bm{r})$ induced by the perturbation $\delta V_\beta(\bm{r'})$ from the approximated Dyson's equation $G=G^0 + G^0VG^0$.

The expression for the susceptibility in Eq. \eqref{eq:chiGG} can easily be extended to a system with several sublattice degrees of freedom, e.g. BLG. In a similar definition for the \textit{magnetic susceptibility} in the spin-density functional formalism, one can define the change in the density as \cite{Kubler} $\delta n_{\alpha\beta}(\bm{r})=n_{\alpha\beta}(\bm{r})-n^0_{\alpha\beta}(\bm{r}) =\sum_{\alpha'\beta'}\int d\bm{r'} \chi_{\alpha\beta,\alpha'\beta'}(\bm{r},\bm{r'})V_{\alpha'\beta'}(\bm{r'})$ where $\alpha$ or $\beta$ denote the sublattice indices (e.g. $A1$, $B1$, $A2$ and $B2$ for BLG) satisfying the closure relationship $\sum_\nu\int\vert\bm{r},\alpha\rangle\langle\bm{r},\alpha\vert d\bm{r}=1$ with the collective sublattice index $\nu={\alpha, \beta,...}$ and the perturbing potential is defined as $V_{\alpha\beta}(\bm{r},\bm{r'})=V_{\alpha\beta}(\bm{r})\delta(\bm{r}-\bm{r'})$. Following similar steps, we find the generalized susceptibility as $\chi_{\alpha\beta,\alpha'\beta'}(\bm{r},\bm{r'})=
\frac{-2}{\pi} \int^{E_{\rm F}}_{-\infty} dE \ \Im m[G^0_{\alpha \alpha'} (\bm{r}, \bm{r'}, E)   G^0_{\beta \beta'} (\bm{r'},\bm{r}, E)]$, and If we restrict the response only to the diagonal external potential, the susceptibility in terms of the diagonal density matrix is given by $\chi_{\alpha \beta} (\bm{r},\bm{r'}) \equiv\delta n_\alpha(\bm{r}) / \delta V_\beta(\bm{r'})$, which finally yields
\begin{eqnarray}
\chi_{\alpha \beta}(\bm{r},\bm{r'})=
- \frac{2}{\pi} \int^{E_{\rm F}}_{-\infty} dE \ \Im m[G^0_{\alpha \beta} (\bm{r}, \bm{r'}, E) G^0_{\beta \alpha} (\bm{r'},\bm{r}, E)].
\label{eq:chiGG2}
\end{eqnarray}
Based on Eq. \eqref{eq:chiGG2}, for two magnetic moments one located at $(\alpha,0)$ and the other at $(\beta,\bm{R})$, we can re-write Eq. \eqref{eq:RKKYJ} for sublattice components of the exchange integral as
\begin{eqnarray}
J_{\alpha \beta} (\bm{R}) = \frac{\lambda^2  }{4} \chi_{\alpha \beta} (0,\bm{R}).
\label{eq:RKKYJ2}
\end{eqnarray}
Knowing the real-space GFs, Eqs. \eqref{eq:chiGG2} and \eqref{eq:RKKYJ2} are the central formulas for calculating different sublattice components of the RKKY interaction in BLG.

\subsection{Green's functions for the effective four-band Hamiltonian: Unbiased Case}

In the absence of the perpendicular electric field ($V=0$), the unperturbed GF in momentum space corresponding to the four-band Hamiltonian ${\cal H}_0$ of Eq. \eqref{eq:H04bnd} is represented by
\begin{widetext}
\begin{align}
G^0(\bm k, \varepsilon)=\frac{1}{\Delta}\,\begin{pmatrix}
\varepsilon\left(\varepsilon^2-f(\bm k)f^*(\bm k)\right) & f(\bm k)\left(\varepsilon^2-f(\bm k)f^*(\bm k)\right) & \varepsilon^2t_{\perp}&f^*(\bm k)\varepsilon t_{\perp}\\
f^*(\bm k)\left(\varepsilon^2-f(\bm k)f^*(\bm k)\right) & \varepsilon\left(\varepsilon^2-f(\bm k)f^*(\bm k)-t_{\perp}^2\right)&f^*(\bm k)\varepsilon t_{\perp}&f^{*2}(\bm k)t_{\perp}\\
\varepsilon^2t_{\perp}&f(\bm k) \varepsilon t_{\perp}& \varepsilon\left(\varepsilon^2-f(\bm k)f^*(\bm k)\right)& f^*(\bm k)\left(\varepsilon^2-f(\bm k)f^*(\bm k)\right) \\
f(\bm k) \varepsilon t_{\perp}&-f^{2}(\bm k)t_{\perp}& f(\bm k)\left(\varepsilon^2-f(\bm k)f^*(\bm k)\right) & \varepsilon\left(\varepsilon^2-f(\bm k)f^*(\bm k)-t_{\perp}^2\right)\end{pmatrix},
\label{eq:G04bnd}
\end{align}
\end{widetext}
where $\varepsilon=E+i0^+$ and $\Delta=\left(\varepsilon^2-f(\bm k)f^*(\bm k)-\varepsilon t_{\perp}\right)\left(\varepsilon^2-f(\bm k)f^*(\bm k)+\varepsilon t_{\perp}\right)$, zeros of which obtain the dispersion relation $E(k)=\pm t_{\perp}/2\,\pm\sqrt{f(\bm k)f^*(\bm k)+(t_{\perp}/2\,)^2}$ leading to the celebrated Mexican-hat band structure of BLG. For sublattices, $A_1$, $B_1$, $A_2$, and $B_2$, the GF expression in Eq. \eqref{eq:G04bnd} has only six independent matrix elements, namely: $ G^0_{A_1A_1}$, $G^0_{A_1A_2}$, $G^0_{B_1B_1}$, $G^0_{A_1B_1}$, $G^0_{A_1B_2}$ and $G^0_{B_1B_2}$. The corresponding matrix elements of the real-space GF are obtained from the Fourier transformations of $G^0_{\alpha \beta}(\bm k, \varepsilon)$ elements, viz., $G^0_{\alpha\beta}(\bm r, \bm r', \varepsilon)=\Omega_{\text{BZ}}^{-1}\,\int e^{i\bm k \cdot (\bm r-\bm r')} G^0_{\alpha\beta}(\bm k, \varepsilon) \,d \bm k$, where $\Omega_{\text{BZ}}=(2\pi)^2/\Omega_{\text{cell}}$ is the area of the first BZ with the area of the
unit cell of SLG to be $\Omega_{\text{cell}}=3\sqrt{3}a^2/2$.

Within the Dirac-cones approximation, the Fourier relationship simplifies and the general real-space GF element connecting the points $(\alpha,0)$ and $(\beta,\bm R)$ is given by
\begin{eqnarray}
G^0_{\alpha\beta} (0,\bm R,\varepsilon)=&\frac{1}{\Omega_{\text{BZ}}}\int d{\bm q} \ e^{-i{\bm q}\cdot{\bm R}}[e^{-i{\bm K}\cdot{\bm R}} G^0_{\alpha\beta}({\bm q+\bm K},\varepsilon) \nonumber\\
&+e^{-i{\bm K'} \cdot {\bm R}} G^0_{\alpha\beta}(\bm q+\bm K',\varepsilon)],
\label{eq:G0Fourier}
\end{eqnarray}
from which the replacement $\bm R \rightarrow -\bm R$ yields the expression for $G^0_{\beta \alpha} (\bm R,0,\varepsilon)$. The details of integrations for the Fourier integral in Eq. \eqref{eq:G0Fourier} are very similar to what has been reported in Ref. [\onlinecite{MSashi1}] and we will not repeat them here. Here, we just report the final results for the six matrix elements. They are
\begin{widetext}
\begin{align}
G^0_{A_1A_1}(0,\bm R,\varepsilon)&=\zeta \varepsilon (e^{-i\bm K \cdot \bm R}+e^{-i\bm K' \cdot \bm R}) [K_0(\sqrt{-\alpha^2}R)+K_0(\sqrt{-\beta^2}R)]\nonumber\\
G^0_{A_1B_1}(0,\bm R,\varepsilon)&=i\zeta v_F (e^{-i\bm K \cdot \bm R+i\theta_R}-e^{-i\bm K' \cdot \bm R-i\theta_R})[\sqrt{-\alpha^2}K_1(\sqrt{-\alpha^2}R)+\sqrt{-\beta^2}K_1(\sqrt{-\beta^2}R)]\nonumber\\
G^0_{A_1A_2}(0,\bm R,\varepsilon)&=\zeta \varepsilon (e^{-i\bm K \cdot \bm R}+e^{-i\bm K' \cdot \bm R})[K_0(\sqrt{-\alpha^2}R)-K_0(\sqrt{-\beta^2}R)]\nonumber\\
G^0_{A_1B_2}(0,\bm R,\varepsilon)&=-i\zeta v_F (e^{-i\bm K \cdot \bm R-i\theta_R}-e^{-i\bm K' \cdot \bm R+i\theta_R})[\sqrt{-\alpha^2}K_1(\sqrt{-\alpha^2}R)-\sqrt{-\beta^2}K_1(\sqrt{-\beta^2}R)]\nonumber\\
G^0_{B_1B_1}(0,\bm R,\varepsilon)&=\zeta (e^{-i\bm K \cdot \bm R}+e^{-i\bm K' \cdot \bm R})[(\varepsilon-t_{\perp})K_0(\sqrt{-\alpha^2}R)+(\varepsilon+t_{\perp})K_0(\sqrt{-\beta^2}R)]\nonumber\\
G^0_{B_1B_2}(0,\bm R,\varepsilon)&=\frac{\zeta v_F^2}{\varepsilon}\, (e^{-i\bm K \cdot \bm R-2i\theta_R}+e^{-i\bm K' \cdot \bm R+2i\theta_R})[\alpha^2 K_2(\sqrt{-\alpha^2}R)-\beta^2 K_2(\sqrt{-\beta^2}R)],
\label{eq:G04}
\end{align}
\end{widetext}
where $\theta_R=\tan^{-1}(y/x)$ is the polar angle of the direction of $\bm R$ with respect to the $x$-axis chosen to be along the direction of $\bm K-\bm K'$, $\zeta=-\pi v_F^{-2}\Omega_{\text{BZ}}^{-1}$, $\alpha^2=v_F^{-2}(\varepsilon^2-\varepsilon t_{\perp})$,  $\beta^2=v_F^{-2}(\varepsilon^2+\varepsilon t_{\perp})$ and $K_\mu(x)$ is the modified Bessel function of the second kind and order of $\mu=0, 1, 2$.

\subsection{Green's functions for the effective two-band Hamiltonian}

Using the Hamiltonian in Eq. \eqref{eq:H02bnd}, the momentum-space matrix representation of the retarded GF defined as $G^0(\varepsilon)=(\varepsilon-H_0)^{-1}$, is given by
\begin{eqnarray}
G^0(\bm q,\varepsilon)=\frac{1}{\Delta'}\,\begin{pmatrix}\varepsilon+V/2&-q^2e^{-2is\theta_q}/2m\,\\-q^2e^{2is\theta_q}/2m\,&\varepsilon-V/2\end{pmatrix},
\label{eq:G0mom}
\end{eqnarray}
where $\Delta'=\varepsilon^2-\varepsilon^2_q$ with the band energy dispersion $\varepsilon_q=\pm \sqrt{\frac{q^4}{4m^2}+\frac{V^2}{4}}$. It should be noticed that the retarded GF in Eq. \eqref{eq:G0mom} has two independent terms, namely $G^0_{B_1B_1}(\bm q,\varepsilon)$ and $G^{0}_{B_1B_2}(\bm q,\varepsilon)$ since $G^0_{B_2B_1}(\bm q,\varepsilon)=G^{0^*}_{B_1B_2}(\bm q,\varepsilon)$. Furthermore, $G^0_{B_2B_2}(\bm q,\varepsilon)$ can be obtained from $G^0_{B_1B_1}(\bm q,\varepsilon)$ by replacing $V \rightarrow -V$.

\textit{Points on the same layer}-- Similar to the previous section, we use Eq. \eqref{eq:G0Fourier} to find the corresponding real-space GFs. The Fourier integral can be evaluated in two ways. As the first method, we can plug the expression for $G^0_{B_1B_1}(\bm q,\varepsilon)$ given in Eq. \eqref{eq:G0mom} into Eq. \eqref{eq:G0Fourier} and integrate. After some algebra, we finally obtain the GF in terms of the Meijer G-function as
\begin{eqnarray}
&G^0_{B_1B_1}(0,\bm R,\varepsilon,V)=\frac{2\pi}{\Omega_{\text{BZ}}} (e^{-i \bm K \cdot \bm R}+e^{-i \bm K' \cdot \bm R}) \times \nonumber \\  &\left[-\frac{m(2\varepsilon+V)}{2\sqrt{V^2-4\varepsilon^2}}G_{0,4}^{\,3,0} \!\left( \left. \begin{matrix} 0,\frac{1}{2},\frac{1}{2},0 \end{matrix} \; \right| \, \frac{m^2R^4}{256}(V^2-4\varepsilon^2)  \right)\right],
\label{eq:G0B1B1_Meijer}
\end{eqnarray}
where the function $G_{p,q}^{m,n}$ in the bracket is the Meijer G-function \cite{MeijerG}.

As for the second method, we re-write  the corresponding momentum-space GF as $G^0_{B_1B_1}(\bm q,\varepsilon)={m^2 (2\varepsilon+V)}\sum_{i=1,2} {\xi}^{-1}(\xi+(-1)^{i}q^2)^{-1}$ where $\xi=\sqrt{4m^2\varepsilon^2-m^2V^2}$. After the integrations, the GF reads
\begin{eqnarray}
&G^0_{B_1B_1}(0,\bm R,\varepsilon,V)=\frac{2\pi}{\Omega_{\text{BZ}}} (e^{-i \bm K \cdot \bm R}+e^{-i \bm K' \cdot \bm R}) \times \nonumber \\
&\frac{m^2 (2\varepsilon+V)}{\xi}\left[K_0(\sqrt{\xi}R)-K_0(\sqrt{-\xi}R)\right],
\label{eq:G0B1B1_Bessel}
\end{eqnarray}
We emphasize that $\xi$ is a c-number depending on the range of the energy. Although both of the expressions in Eqs. \eqref{eq:G0B1B1_Meijer} and \eqref{eq:G0B1B1_Bessel} are equivalent, we will be using the one in terms of the modified Bessel function, which will be practical for numerical analysis.

\textit{Points on different layers}-- In this case, we write $G^0_{B_1B_2}(\bm q,\varepsilon)$ as $me^{2is\theta_q}\sum_{i=1,2}(-1)^i(\xi+(-1)^iq^2)^{-1}$. Following the steps elaborated for the same layer, we obtain two equivalent expressions for the real-space GF corresponding to this case in terms of both Meijer G- and modified Bessel functions as
\begin{widetext}
\begin{eqnarray}
G^0_{B_1B_2}(0,\bm R,\varepsilon,V)&=-\frac{m\pi}{\Omega_{\text{BZ}}} \left[e^{-i (\bm K \cdot \bm R+2\theta_R)}+e^{-i (\bm K' \cdot \bm R-2\theta_R)}\right] G_{0,4}^{\,3,0} \!\left( \left. \begin{matrix} 0,\frac{1}{2},1,-\frac{1}{2} \end{matrix} \; \right| \, \frac{m^2R^4}{256}(V^2-4\varepsilon^2)  \right) \nonumber\\
&=\frac{2m\pi}{\Omega_{\text{BZ}}}  \left[e^{-i (\bm K \cdot \bm R+2\theta_R)}+e^{-i (\bm K' \cdot \bm R-2\theta_R)}\right] \left[K_2(\sqrt{\xi}R)+ K_2(\sqrt{-\xi}R)\right]~.
\label{eq:G0B1B2}
\end{eqnarray}
\end{widetext}

\section{RKKY Interaction from the two-band Model}

In this section, for the moments on the same layer, we use the expressions for the real-space GFs, $G^0_{B_1B_1}(0,\bm R,\varepsilon,V)$ and $G^0_{B_1B_1}(\bm R,0,\varepsilon,V)$ from Eq. \eqref{eq:G0B1B1_Bessel} and then obtain the RKKY interaction $J_{B_1B_1}(\bm R)$ using Eqs. \eqref{eq:chiGG2} and \eqref{eq:RKKYJ2}.

\subsection{Moments on the same layer: $J_{B_1B_1}(\bm R)$}

\textit{General case}-- Using Eq. \eqref{eq:G0B1B1_Bessel} and substituting both $G^0_{B_1B_1}(0,\bm R,\varepsilon,V)$ and $G^0_{B_1B_1}(\bm R,0,\varepsilon,V)$ in Eq. \eqref{eq:chiGG2}, the corresponding susceptibility reads
\begin{eqnarray}
\chi_{B_1B_1}(0,\bm R)=\frac{-16\pi m^2}{\Omega_{\text{BZ}}^2}\Phi_{B_1B_1}I_1(V,R,E_F)
\label{eq:chiB1B1}
\end{eqnarray}
where $\Phi_{B_1B_1}=1+\cos\left[(\bm K-\bm K')\cdot \bm R\right]$ and the integral $I_1$ is given by
\begin{eqnarray}
I_1(V,R,E_F)=&\Im m \int^{E_F}_{-\infty} dE \ \left(\frac{2\varepsilon+V}{\sqrt{4\varepsilon^2-V^2}}\right)^2 \times \nonumber \\
&\left[K_0(\sqrt{\xi}R)-K_0(\sqrt{-\xi}R)\right]^2~.
\label{eq:I1B1B1}
\end{eqnarray}

$I_1$ may not be analytically evaluated for arbitrary values of the gate voltage, $V$ and Fermi energy $E_F$. However, for the special case of unbiased system, $V=0$ one can manage to find the analytical expression for $I_1$ as elaborated in the following subsection.

\textit{Special case: Unbiased BLG}-- Now, we split the integral in Eq. \eqref{eq:I1B1B1} into two parts, viz., $ \int_{-\infty}^{E_F}= \int_{-\infty}^{0}+ \int_{0}^{E_F}$, where the first term accounts for the valance electrons (undoped case) and the second for the conduction electrons (doped case). Let's denote the first integral by $I_0$. Special care must be taken to consider the principal value of the complex square roots of the complex variables. For instance, $\xi=2m\sqrt{\varepsilon^2}=\pm2m\varepsilon$ for $E>0$ and $E<0$, respectively. Then, we introduce new variables $y=\pm 2mER^2$ accordingly for both integrals such that $y>0$ in each and express the modified Bessel function with complex argument in terms of the Hankel functions and then Bessel functions of first and second kind using $K_\nu(z)=2^{-1}i \pi e^{i \pi \nu/2}H_\nu^{(1)}(z e^{i \pi/2})=-2^{-1}i \pi e^{-i \pi \nu/2}H_\nu^{(2)}(ze^{-i \pi/2})$ with $H_\nu^{(1,2)}(z)=J_\nu(z) \pm i \ Y_\nu(z)$. In particular, we use $K_0(\sqrt{y\pm
i 0^+})=K_0(\sqrt{y})$ and $K_0(\sqrt{-y \pm i 0^+})=(-\pi/2)\left(Y_0(\sqrt{y})\pm i J_0(\sqrt{y})\right)$.

After some algebra, $I_0$ simplifies to
\begin{eqnarray}
I_0=\frac{\pi}{2mR^2}[&\int_{0}^{\infty} dy J_0(\sqrt{y}) K_0(\sqrt{y})+ \nonumber \\
\frac{\pi}{2}&\int_{0}^{\infty} dy J_0(\sqrt{y}) Y_0(\sqrt{y})].
\label{eq:I0}
\end{eqnarray}
Both integrals in Eq. \eqref{eq:I0} are diverging; however, after using some regulatory cut-off functions \cite{MSashi1} they are evaluated to one and zero, respectively, which yields $I_0=\pi(2mR^2)^{-1}$. Plugging this result into Eqs. \eqref{eq:chiB1B1} and \eqref{eq:RKKYJ} immediately gives the RKKY interaction for the unbiased and undoped BLG as
 \begin{eqnarray}
J^0_{B_1B_1}(\bm R)=-C\frac{1+\cos\left[(\bm K-\bm K')\cdot \bm R\right]}{(R/a)^2},
 \label{eq:JB1B1V0EF0}
 \end{eqnarray}
where $C=3  \lambda^2 (16\pi^2 t^2)^{-2}t_\perp$ is a positive parameter, which means that $J^0_{B_1B_1}(\bm R)$ represents an FM interaction between the moments. The power-law $R^{-2}$ decay of the RKKY interaction in Eq. \eqref{eq:JB1B1V0EF0} clearly shows that the undoped and unbiased BLG behaves like an ordinary 2D electron gas, the result that have also been reported in other studies \cite{Hwang, Killi, Kogan, Jiang}.

As for the general doped case, after similar simplifications, we obtain
\begin{widetext}
\begin{eqnarray}\label{eq:i1}
I_1(V=0,R,x_F)=\frac{\pi}{2mR^2}\left[1-\int_{0}^{x_F} dy J_0(\sqrt{y}) K_0(\sqrt{y})-\frac{\pi}{2}\int_{0}^{x_F} dy J_0(\sqrt{y})Y_0(\sqrt{y})\right],
\label{eq:i1}
\end{eqnarray}
\end{widetext}
where $x_F=2mE_FR^2=k_F^2R^2$. Both integrals in Eq. \eqref{eq:i1} can be expressed  \cite{MeijerG} in terms of the Meijer G-functions, viz., $\int_{0}^{x_F} dy J_0(\sqrt{y}) K_0(\sqrt{y})=8^{-1}\pi^{-1/2} x_F G_{1,5}^{\,3,1} \!\left( \left. \begin{matrix} \frac{1}{2} \\ 0,0,\frac{1}{2},-\frac{1}{2},0 \end{matrix} \; \right| \, x_F^2/64 \right)$ and similarly, $\int_{0}^{x_F} dy J_0(\sqrt{y}) Y_0(\sqrt{y})=-\pi^{-1/2} G_{1,3}^{\,2,0} \!\left( \left. \begin{matrix} \frac{3}{2} \\ 1,1,0 \end{matrix} \; \right| \, x_F \right)$~.

We find the asymptotic expansion of the Meijer G-functions  \cite{MeijerG} in Eq. \eqref{eq:i1} and eventually obtain the RKKY interaction for the large distances as
\begin{widetext}
\begin{eqnarray}\label{eq:i1a}
\lim _{k_FR \rightarrow \infty} I_1(V=0,R,k_F)=\frac{\pi}{2mR^2}\left[\sqrt{2}e^{-k_FR}\cos{(k_FR)}+\frac{1}{2}\sin{(2k_FR)}-\frac{\cos{(2k_FR)}}{8k_FR}\right]~.
\end{eqnarray}
\end{widetext}

The peculiar feature of the interaction at this limit is the exponential decay along with the power-law decay. However, the exponential term does not survive for very large distances. The reason for such an exponential decay can probably be explained based on the chiral characteristics of the quasiparticle in BLG and the fact that the forward scattering is forbidden in the system.

\subsection{Moments on different layers: $J_{B_1B_2}(\bm R)$}

\textit{General case}-- In this section, we consider the situation in which the magnetic moments are located on different layers. Using Eqs. \eqref{eq:G0B1B2} and \eqref{eq:chiGG2}, the corresponding susceptibility, $\chi_{B_1B_2}(0,\bm R)$ is given by
\begin{eqnarray}
\chi_{B_1B_2}(0,\bm R)=\frac{-16 \pi m^2}{\Omega_{\text{BZ}}^2} \ \Phi_{B_1B_2} \ I_2(V,R,E_F),
\label{eq:chiB1B2}
\end{eqnarray}
where $\Phi_{B_1B_2}=1+\cos\left[(\bm K-\bm K') \cdot \bm R+4 \theta_R \right]$ and the remaining integral is given by
\begin{eqnarray}\label{eq:i2vef}
I_2(V,R,E_F)=\int^{E_F}_{-\infty} dE \ \Im m \left[K_2(\sqrt{\xi}R)+K_2(\sqrt{-\xi}R)\right]^2~.
\end{eqnarray}
Similar to the case of the moments on the same layer, we can find the analytical expression for the RKKY interaction for the unbiased BLG.

\textit{Special case: Unbiased BLG}--  Here, again we split the integral of Eq. \eqref{eq:i2vef} into the undoped and doped parts and perform the same type of calculations as explained before. In particular, we use $K_2(\sqrt{y\pm i 0^+})=K_2(\sqrt{y})$ and $K_2(\sqrt{-y\pm i 0^+})=(\pi/2)\left(Y_2(\sqrt{y})\pm i J_2(\sqrt{y})\right)$. Therefore, the integral corresponding to the undoped part is denoted by $I'_0$ and it reads
\begin{eqnarray}
I'_0=\frac{\pi}{2mR^2}[&\int_{0}^{\infty} dy J_2(\sqrt{y}) K_2(\sqrt{y})+ \nonumber \\
\frac{\pi}{2}&\int_{0}^{\infty} dy J_2(\sqrt{y}) Y_2(\sqrt{y})].
\label{eq:Ip0}
\end{eqnarray}
Resorting to cut-off function scheme, both diverging integrals in Eq. \eqref{eq:Ip0} give one and $-4/\pi$, respectively, which yields $I'_0=-\pi(2mR^2)^{-1}$. Finally, using $I'_0$ in Eqs. \eqref{eq:chiB1B2} and \eqref{eq:RKKYJ} we obtain the RKKY interaction for the unbiased and undoped BLG for the moments of different layers as
\begin{eqnarray}
J^0_{B_1B_2}(\bm R)=C\frac{1+\cos\left[(\bm K-\bm K') \cdot \bm R+4 \theta_R \right]}{(R/a)^2}.
\label{eq:JB1B2V0EF0}
\end{eqnarray}
As $C>0$, $J^0_{B_1B_1}(\bm R)$ signifies an AFM interaction between the moments on different layers.

The comparison between $J^0_{B_1B_2}(\bm R)$ and $J^0_{B_1B_1}(\bm R)$ given in Eq. \eqref{eq:JB1B1V0EF0} reveals a very subtle point. Apart from their different oscillatory Dirac-cones factor, which are both bound and positive, $J_{B1B1}$ and $J_{B_1B_2}$ has the same magnitude $C$ and opposite sign. We recall that for the case of undoped SLG, the RKKY interaction for the moments on the opposite sublattices is AFM and its magnitude is three times larger than that of for the same sublattice, namely, $J_{AB}=3J_{AA}$~\cite{Saremi,Brey,MSashi1}, which means the AFM ordering is more favored for SLG. It was reasoned by Saremi~\cite{Saremi} that this commensurate feature of the RKKY interaction must be the case for any bipartite system with particle-hole symmetry. By analogy, our results for $J_{B_1B_1}$ and $J_{B_1B_2}$ may be interpreted as the signature of the bipartite nature of the system and the particle-hole symmetry present in the effective two-band Hamiltonian $H_0$ in Eq.~\eqref{eq:H02bnd}.
Although it appears that the
unbiased BLG bears the same symmetry, an attempt to prove the theorem particularly for this system will be insightful. To our knowledge such proof does not exist.

Similarly, we can find the analytical expressions of the interaction for the doped case. Following same steps as discussed previously, Eq.~\eqref{eq:i2vef} simplifies to
\begin{widetext}
\begin{eqnarray}
I_2(V=0,R,x_F)=-\frac{\pi}{2mR^2}\left[1+\int_{0}^{x_F} dy J_2(\sqrt{y}) K_2(\sqrt{y})+\frac{\pi}{2}\int_{0}^{x_F} dy J_2(\sqrt{y})Y_2(\sqrt{y})\right],
\label{eq:I2B1B2V0}
\end{eqnarray}
\end{widetext}
where the first and second integrals in Eq.~\eqref{eq:I2B1B2V0} are evaluated as $\int_{0}^{x_F} dy J_2(\sqrt{y}) K_2(\sqrt{y})=8^{-1} \pi^{-1/2} x_F G_{1,5}^{\,3,1} \!\left( \left. \begin{matrix} \frac{1}{2} \\ 0,\frac{1}{2},1,-1,-\frac{1}{2} \end{matrix} \; \right| \, x_F^2/64 \right)$ and $\int_{0}^{x_F} dy J_2(\sqrt{y}) Y_2(\sqrt{y})=-\pi^{-1/2} G_{2,4}^{\,2,1} \!\left( \left. \begin{matrix} 1,\frac{3}{2} \\ 1,3,-1,0 \end{matrix} \; \right| \, x_F \right)$, respectively. The long-distance expression is given by
\begin{widetext}
\begin{eqnarray}
\lim _{k_FR \rightarrow \infty} I_2(V=0,R,k_F)=-\frac{\pi}{2mR^2}\left[\sqrt{2}e^{-k_FR}\cos{(k_FR)}-\frac{1}{2}\sin{(2k_FR)}-\frac{15\cos{(2k_FR)}}{8k_FR}\right].
\label{eq:I2B1B2V0Asy}
\end{eqnarray}
\end{widetext}

Comparing Eq.~\eqref{eq:i1a} and \eqref{eq:I2B1B2V0Asy}, we note that the long-range functional form of the RKKY interaction in unbiased BLG for the moments on different layers is the same as for those on the same layer.

\section{RKKY Interaction from four-band continuum model: Unbiased BLG}

In this section, we report the expressions for the RKKY interaction for the unbiased and doped BLG using the four-band model. For the biased case the same analysis can be made, which will not be presented here.

Plugging the GFs from Eq. \eqref{eq:G04} into Eq. \eqref{eq:chiGG2}, the corresponding susceptibilities are given by
\begin{widetext}
\begin{align}
\chi_{A_1A_{1(2)}}(0,\bm R)&= \Lambda \Phi_{A_1A_{1(2)}} \int_{-\infty}^{x_F} x^2 \Im m\left[K_0(\sqrt{-x^2+\frac{t_{\perp} R}{v_{\rm F}}x}
\pm K_0(\sqrt{-x^2-\frac{t_{\perp} R}{v_{\rm F}}~x})\right]^2 \,\mathrm{d}{x}\nonumber\\
\chi_{A_1B_{1(2)}}(0,\bm R)&= \Lambda \Phi_{A_1B_{1(2)}} \int_{-\infty}^{x_F} \Im m\left[\sqrt{-x^2+\frac{t_{\perp} R}{v_{\rm F}}\,x}K_1(\sqrt{-x^2+\frac{t_{\perp} R}{v_{\rm F}}\,x})\pm \sqrt{-x^2-\frac{t_{\perp} R}{v_{\rm F}}\,x}K_1(\sqrt{-x^2-\frac{t_{\perp} R}{v_{\rm F}}\,x})\right]^2 \,\mathrm{d}{x}
\label{eq:wb0}
\end{align}
\begin{align}
\chi_{B_1B_1}(0,\bm R)&= \Lambda\Phi_{B_1B_1} \int_{-\infty}^{x_F} \Im m\left[(x-\frac{t_{\perp}R}{v_{\rm F}}\,)K_0(\sqrt{-x^2+\frac{t_{\perp} R}{v_{\rm F}}\,x})+(x+\frac{t_{\perp}R}{v_{\rm F}}\,)K_0(\sqrt{-x^2-\frac{t_{\perp} R}{v_{\rm F}}\,x})\right]^2 \,\mathrm{d}{x}
\label{eq:wb1}
\end{align}
\begin{align}
\chi_{B_1B_2}(0,\bm R)&= \Lambda \Phi_{B_1B_2}
\int_{-\infty}^{x_F}\frac{1}{x^2}\,\Im
m\left[(x^2-\frac{t_{\perp}R}{v_{\rm
F}}x\,)K_2(\sqrt{-x^2+\frac{t_{\perp} R}{v_{\rm
F}}\,x})-(x^2+\frac{t_{\perp}R}{v_{\rm
F}}\,)K_2(\sqrt{-x^2-\frac{t_{\perp} R}{v_{\rm F}}\,x})\right]^2
\,\mathrm{d}{x},
\label{eq:wb2}
\end{align}
\end{widetext}
where $x=RE/v_{\rm F}$, $\Lambda =-4\pi \Omega_{\text{BZ}}^{-2} v_{\rm F}^{-1}R^{-3}$ and $x_F=RE_{\rm F}/v_{\rm F}$,
$\Phi_{A_1A_1}=\Phi_{B_1B_1}=\Phi_{A_1A_2}=1+\cos\left[(\bm K-\bm K')\cdot \bm R\right]$, $\Phi_{A_1B_{1(2)}}=1-\cos \left[(\bm K-\bm K')\cdot \bm R \mp 2\theta_R\right]$ and $\Phi_{B_1B_2}=1+\cos \left[(\bm K-\bm K')\cdot \bm R+4\theta_R\right]$.

To calculate the RKKY interaction in the four-band continuum model, Eqs.~(\ref{eq:wb0})-(\ref{eq:wb2}) may be evaluated numerically. The results of Eqs.~(\ref{eq:wb1})-(\ref{eq:wb2}) and those obtained from the two-band model are compared in the following section.

\section{Numerical Result}

In this section, we present our main calculations for the exchange coupling of the RKKY interaction evaluating Eqs.~(\ref{eq:I1B1B1}), (\ref{eq:i2vef}) and (\ref{eq:wb0})-(\ref{eq:wb2}). The general features of the exchange coupling, basically the dependence of the RKKY interaction on the distance $R$ have been numerically studied previously~\cite{Jiang} for the unbiased and undoped BLG. We present, on the other hand, our numerical results of $I_1(V, R, E_{\rm F})$ and $I_2(V, R, E_{\rm F})$ for biased BLG in two different interesting regimes namely doped and undoped graphene where $E_{\rm F}=0$ and $E_{\rm F} \neq 0$, respectively. We provide a comparison between the results obtained within the four-band and the two-band continuum models in unbiased BLG systems and discuss the discrepancy between two models.

\subsection{Unbiased and doped BLG}

In previous Sections, we found the analytical expressions of the RKKY interaction for the unbiased BLG within the two-band model. We showed that regardless of the Dirac-cones oscillatory term represented by $\Phi_{\alpha,\beta}$, the main difference between the interactions for the moments on the same and different layers is due to their sign, which results in FM interaction between impurities on the same layer $B_1B_1$ and AFM interaction in $B_1B_2$ case.
By doping BLG, the strength of the RKKY interaction decreases and a new oscillatory behavior starts. Therefore, the RKKY interaction changes sign as a function of distance.

\begin{figure}
\includegraphics[width=1.0\linewidth]{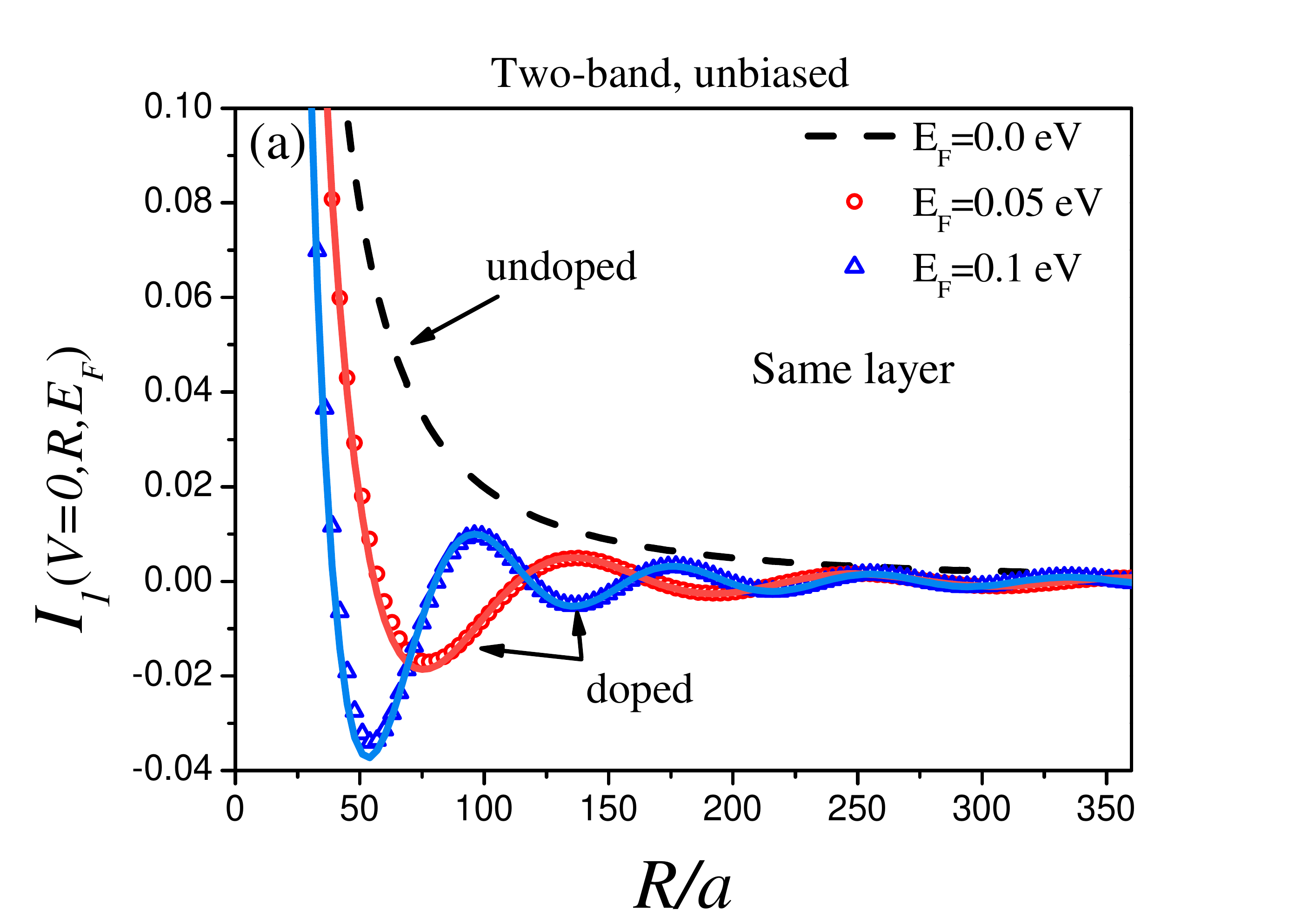}
\includegraphics[width=1.0\linewidth]{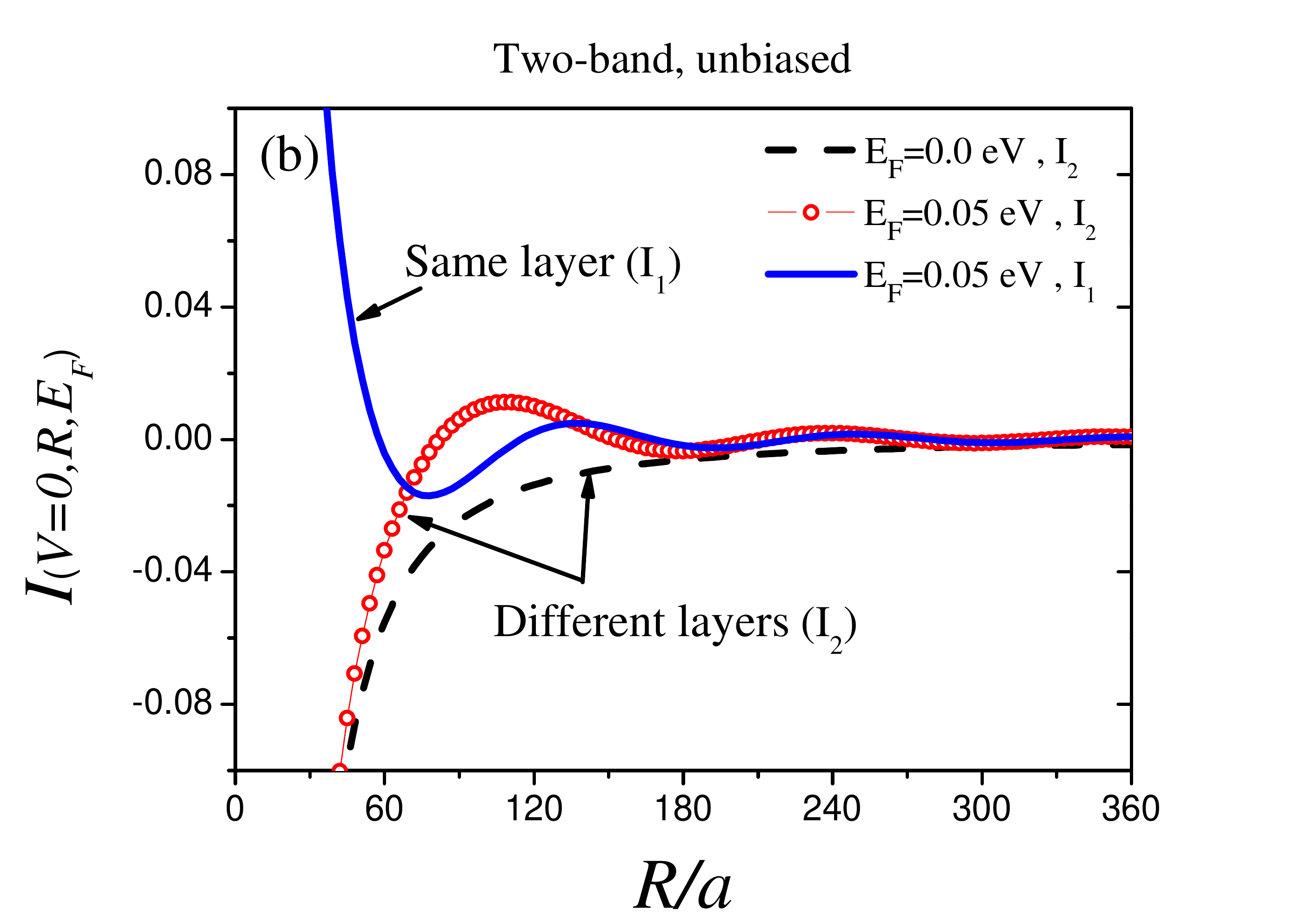}
\caption{ (Color online) (a) $I_1( V=0, R, E_{\rm F})$ as a function of the distance $R$ for different doping values obtained for the two-band model and unbiased BLG using Eq. \eqref{eq:i1}. The function $ R^2 I_1( V=0, R, E_{\rm F}=0)$ is a constant in agreement with that result obtained in Ref.~[\onlinecite{Kogan}]. Solid lines refer to the analytical results of the asymptotic behavior from Eq.~(\ref{eq:i1a}) which are compared to the numerical evaluation of Eq.~(\ref{eq:i1}), plotted as symbols, show their difference at short distances while reaching each other as $R$ increases.
(b) The strength of the interaction for both $B_1B_1$ [Eq. \eqref{eq:i1}] and $B_1B_2$ [Eq. \eqref{eq:I2B1B2V0}] for undoped and doped systems for $E_{\rm
F}=0.05$ eV as a function of the distance.
\label{fig:Graph01}}
\end{figure}

Fig.~\ref{fig:Graph01}(a) shows $I_1( V=0, R, E_{\rm F})$ integral as a function of the distance, $R$ for different doping values. The period of oscillation and the speed of a decay depends strongly on the Fermi Energy. For non-zero $E_{\rm F}$, the integral $I_1$ shows a quite different behavior as it exhibits an oscillatory behavior as a function of $R$ with decreasing amplitude and a period given by $\pi/k_{\rm F}$. We compare the analytical results of Eq.~(\ref{eq:i1a}), plotted as solid lines, with the numerical evaluation of Eq.~(\ref{eq:i1}), plotted as symbols, to show their difference at short distance while reaching each other quite well at large $R$
regions. A comparison between $R$-dependence of the integral $I_1$ and that of $I_2$ for $E_{\rm F}=0.05$ eV in Fig.~\ref{fig:Graph01}(b), shows their difference at short distance while reaching each other approximately as $R$ increases. Similar to $I_1$, at finite $E_{\rm F}$, the integral $I_2$ has an oscillatory behavior as a function of $R$, with decreasing amplitude and a period
given by $\pi/k_{\rm F}$.

\begin{figure}
\includegraphics[width=1.0\linewidth]{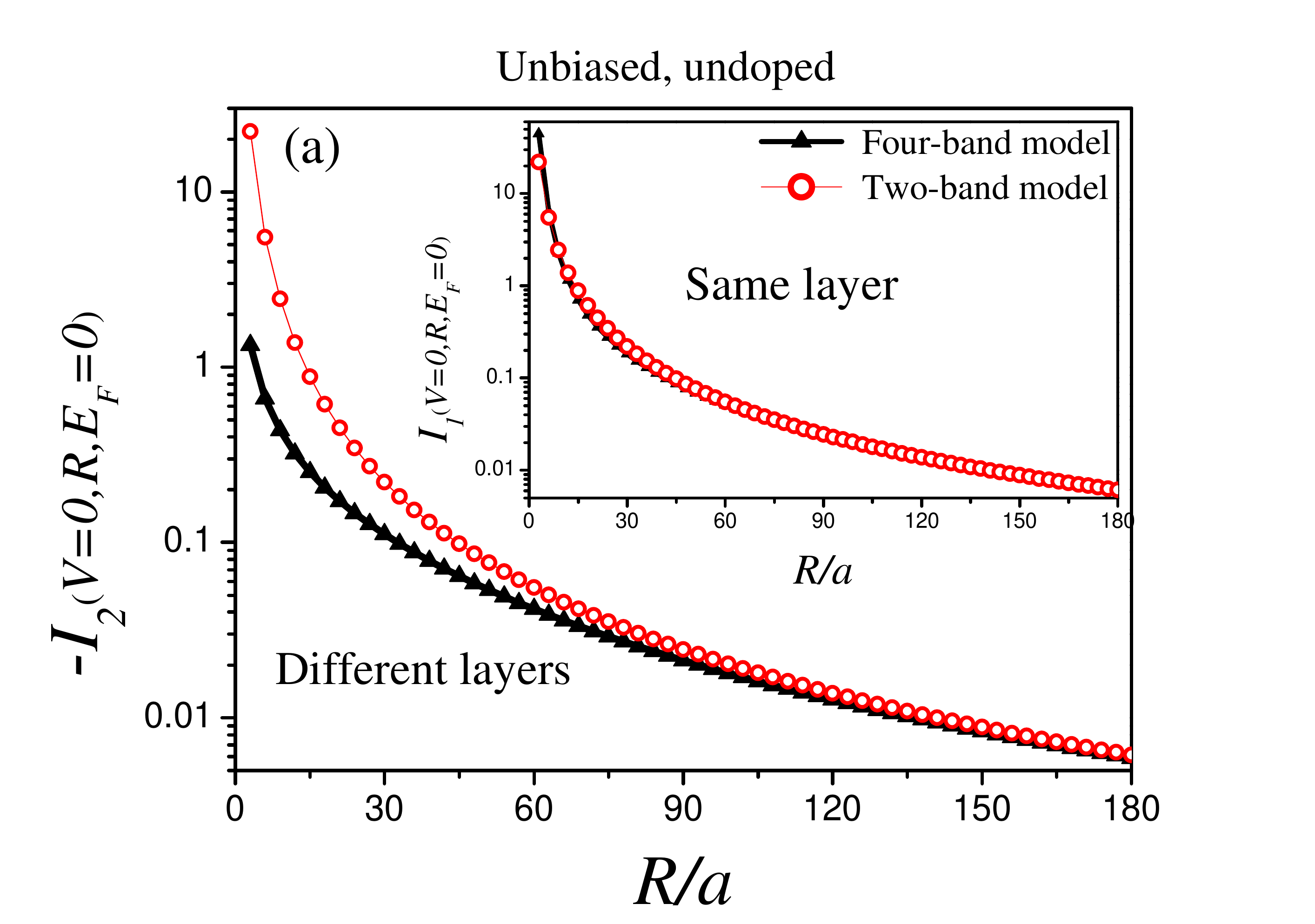}
\includegraphics[width=1.0\linewidth]{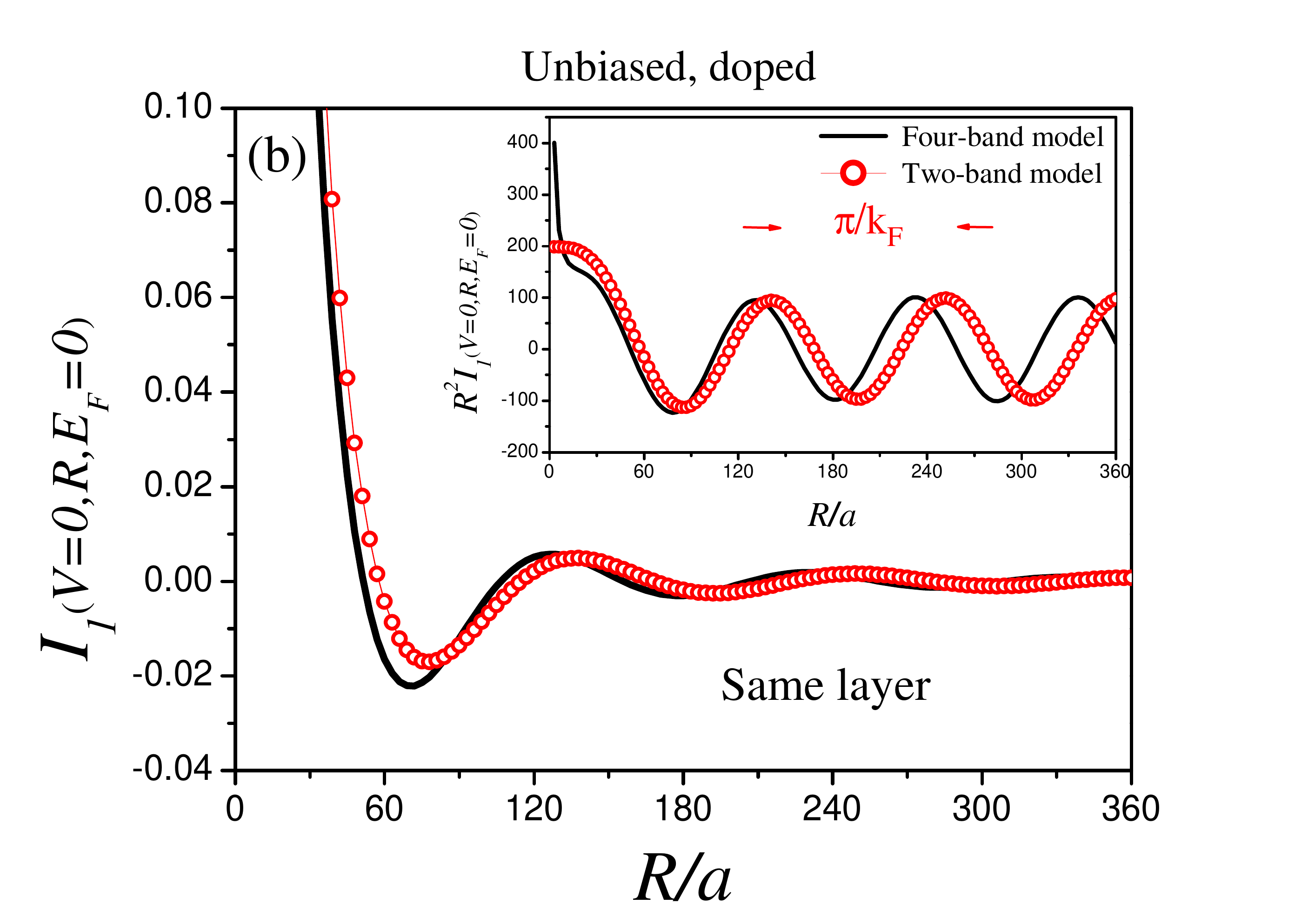}
\includegraphics[width=1.0\linewidth]{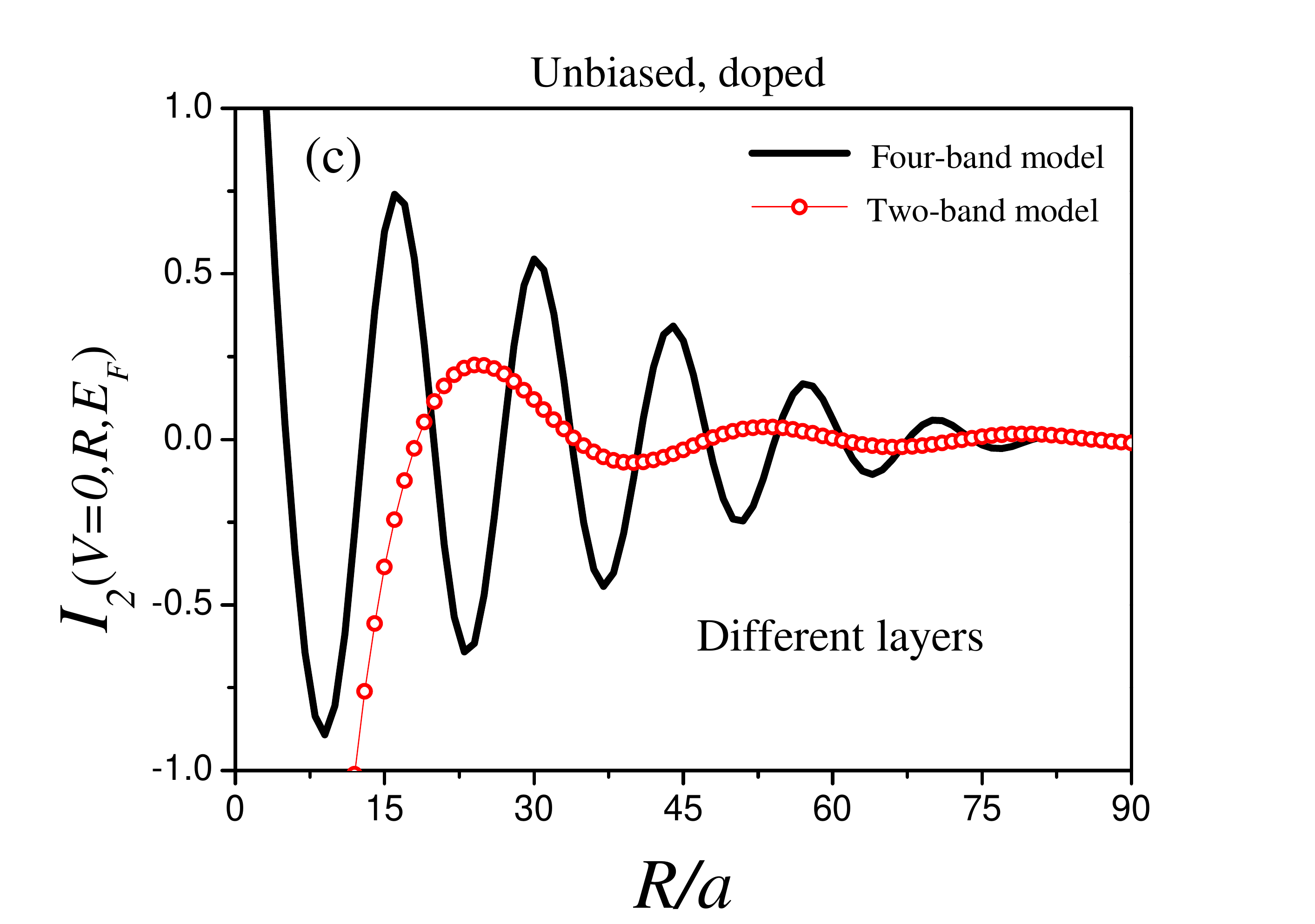}
\caption{(Color online) (a) Comparison between the results of the RKKY interaction that obtained by the two-band model [Eq.~\eqref{eq:I2B1B2V0}] and that calculated by the four-band model [Eq.~\eqref{eq:wb2}] for $I_2$ at $V=E_{\rm F}=0$. There is a discrepancy between two approaches which basically comes from the off-diagonal inter-layer tunneling term $t_{\perp}$. In the inset, a comparison of results between two theories for the case of $I_1$. (b) $I_1( V=0, R, E_{\rm F})$ for $E_{\rm F}=0.05$ eV and (c) $ I_2( V=0, R, E_{\rm F})$ for $E_{\rm F}=1$ eV as a function of the distance $R$ obtained by the two-band model [Eqs.~(\ref{eq:i1}) and (\ref{eq:I2B1B2V0})] and that calculated by the four-band model [Eqs.~(\ref{eq:wb1}) and (\ref{eq:wb2})], respectively. The proper results of the quasiparticle excitation are captured by the four-band model, by increasing the Fermi energy.
\label{fig:4bandvs2band}}
\end{figure}

Fig.~\ref{fig:4bandvs2band} shows a comparison between the RKKY interaction that calculated by the four-band model given by Eqs.~(\ref{eq:wb1}), (\ref{eq:wb2}) and those obtained by two-band model. For an undoped case, it can be seen from Fig.~\ref{fig:4bandvs2band}(a) that two results are matched for FM interaction, ($I_1$) while there is a discrepancy between results at short-range regime which is controlled by value of $t_{\perp}$ for AFM interaction, ($I_2$). This result is very pertinent to the conclusion stated in Ref.~[\onlinecite{borghi_4band}] where the authors show that based on the charge-charge response function calculations, the density-sum and density-difference fluctuations in BLG crossover from those of an unusual massive-chiral single-layer system to those of a weakly coupled bilayer as carrier density, wave vector, and energy increase. Fig.~\ref{fig:4bandvs2band}(b) shows the $I_1$ for unbiased and doped BLG when $E_{\rm F}=0.05$ eV calculated by the four and two band continuum models. It is
clear that the period of the oscillation in the four-band model is different from that of the two-band model. This is because for a certain and small Fermi energy value, the associated Fermi momentum is larger than the value obtained in the two-band model. Note that the electronic dispersion relations in the four-band model (roots of $\Delta$ defined in Eq. \eqref{eq:G04bnd}) may be written as $E_{1,2}(k)=\sqrt{ v^2_{\rm F}k^2+t^2_{\perp}/4} \pm t_{\perp}/2$ and $E_{3,4}(k)=\sqrt{ v^2_{\rm F}k^2+t^2_{\perp}/4} \pm t_{\perp}/2$. Therefore, the period of the oscillations depends on the model.

In the four-band model, depending on the doping level, the Fermi energy can have either one or two intersections with the conduction-band Fermi surfaces. By increasing the Fermi energy, the proper results of the quasiparticle excitation are captured by the four-band model. This point is demonstrated in Fig.~\ref{fig:4bandvs2band}(c) where we show the results of the $I_2$ for $E_{\rm F}=1$ eV, for which the Fermi energy intersects two conduction bands. It is clear that in this case the results obtained by full band are completely different from those calculated by the low-energy excitation method.

\begin{figure}
\includegraphics[width=1.0\linewidth]{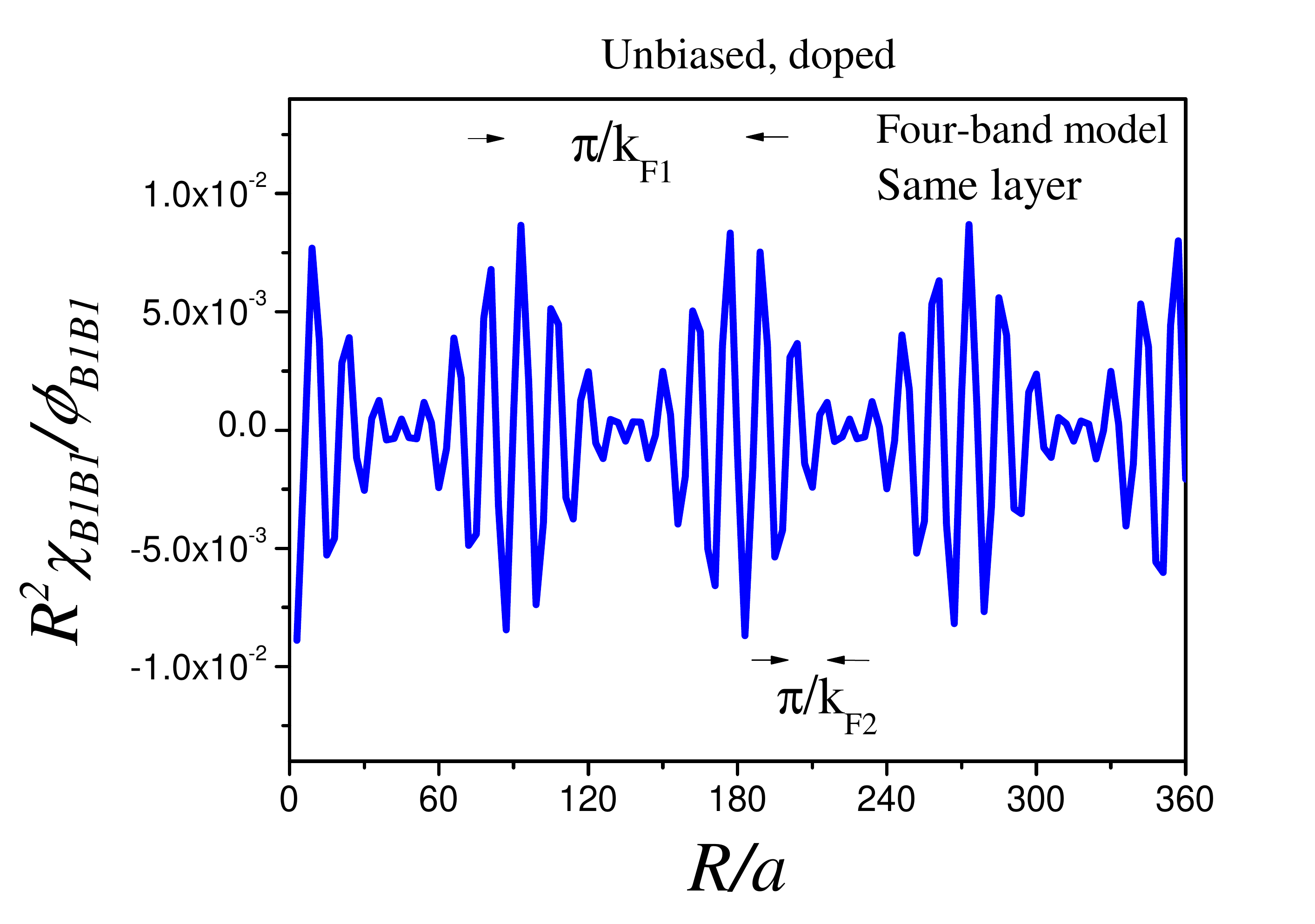}
\caption{(Color online) The susceptibility $\chi_{B_1B_1}$ as a function of the distance between impurities on the same sublattice along armchair direction obtained from the four-band model [Eqs.~\eqref{eq:wb1}]. The existence of two different periods (beating pattern) in doped BLG for certain values of
$E_{\rm F}=1$ eV is clear in this figure.
\label{fig:beat}}
\end{figure}

\begin{figure}
\includegraphics[width=1.0\linewidth]{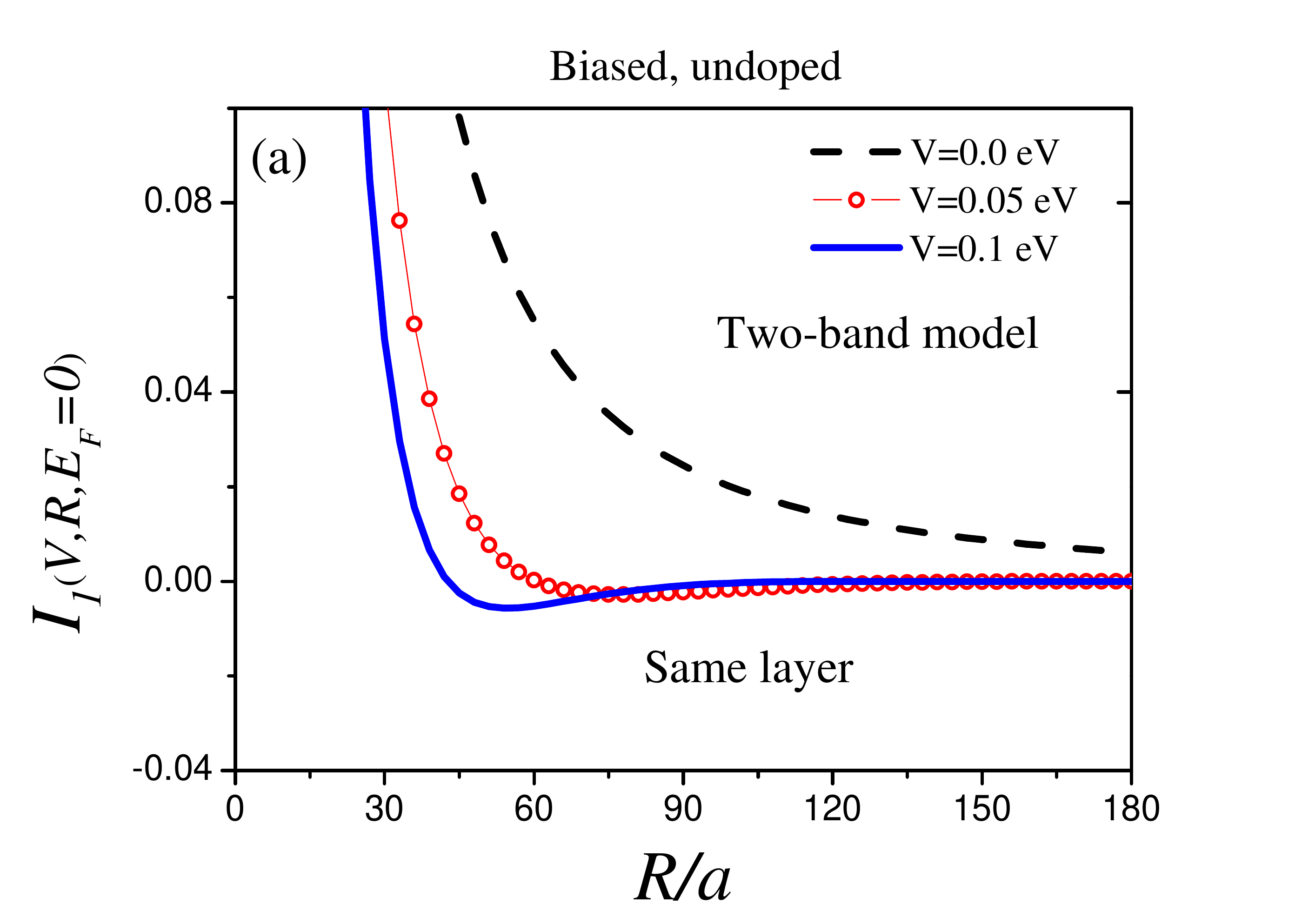}
\includegraphics[width=1.0\linewidth]{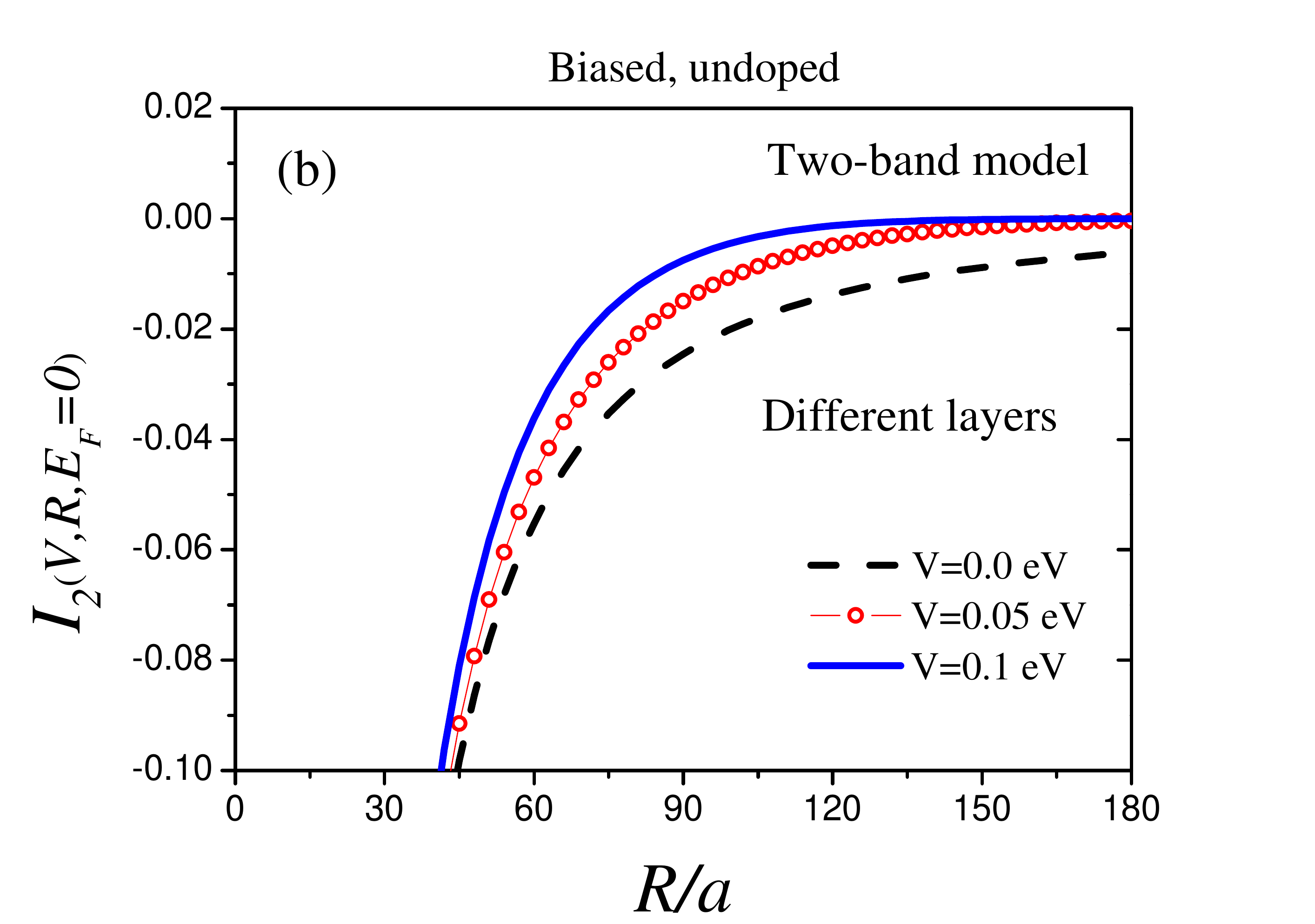}
\caption{ (Color online) (a) The integral $I_1(V, R, E_{\rm F}=0)$ as a function of $R$ for different gate voltages.
The function falls off rapidly and oscillates slightly for finite $V$ values. (b) the same as (a) but for the integral $I_2(V, R, E_{\rm F}=0)$.
\label{fig:Graph03}}
\end{figure}

The intersection of the Fermi energy with the bands denoted by $E_{1(3)}$ creates two Fermi surfaces. Because the RKKY interaction is fundamentally
determined by the geometrical features of the Fermi surface of the host material, a somewhat more complicated behavior of the RKKY coupling for a highly doped BLG can occur. As the result, we observe that oscillations of $\chi_{B_1B_1}$ exhibit a beating pattern with two
characteristic periods associated with the two Fermi momenta defined as $k_{\rm F_{1(2)}}=\sqrt{E^2_{\rm F} \mp E_{\rm F}t_{\perp}}/v_{\rm F}$. Fig.~\ref{fig:beat} shows this beating pattern the RKKY interaction as a function of the distance between impurities.

\begin{figure}
\includegraphics[width=1.0\linewidth]{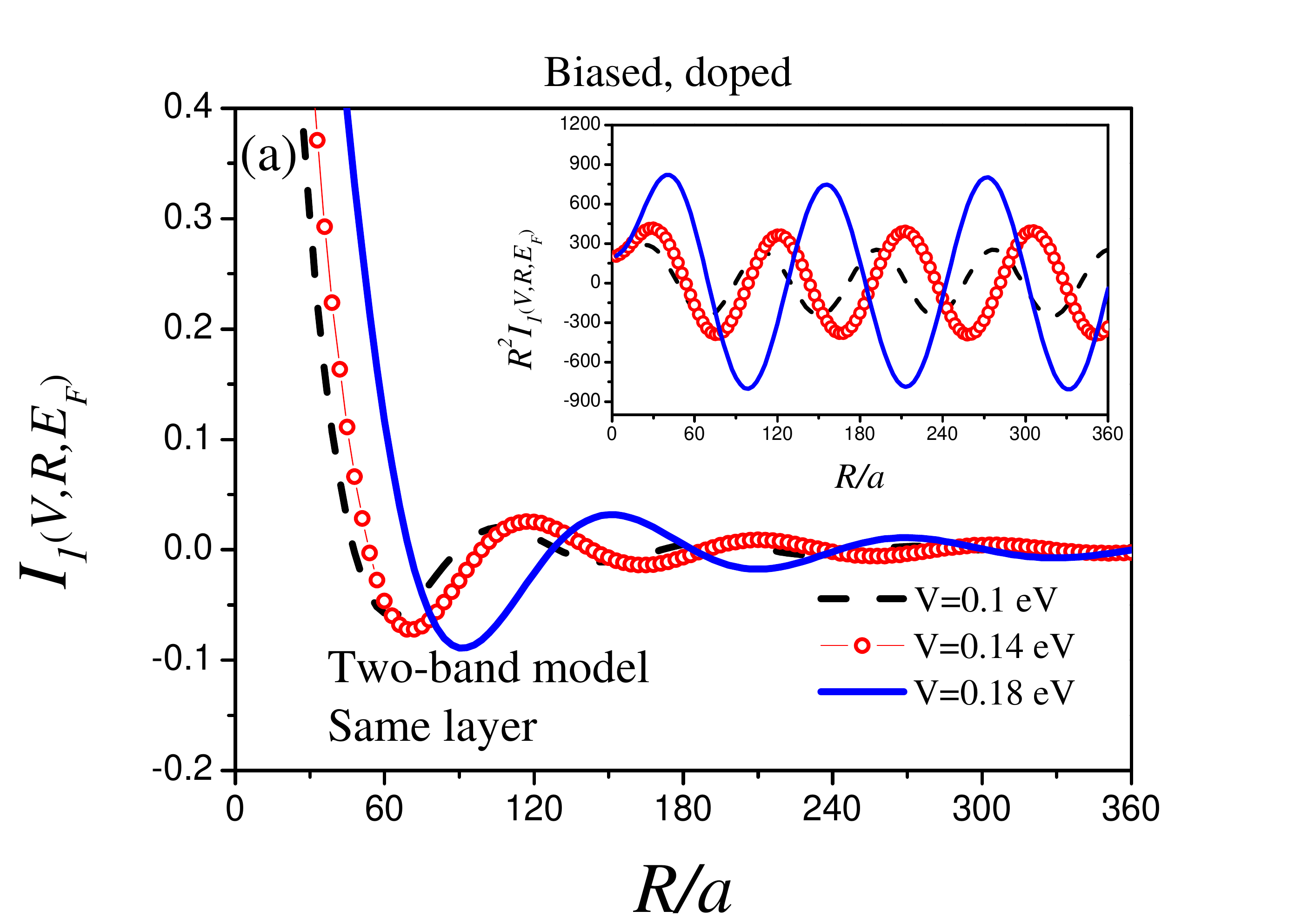}
\includegraphics[width=1.0\linewidth]{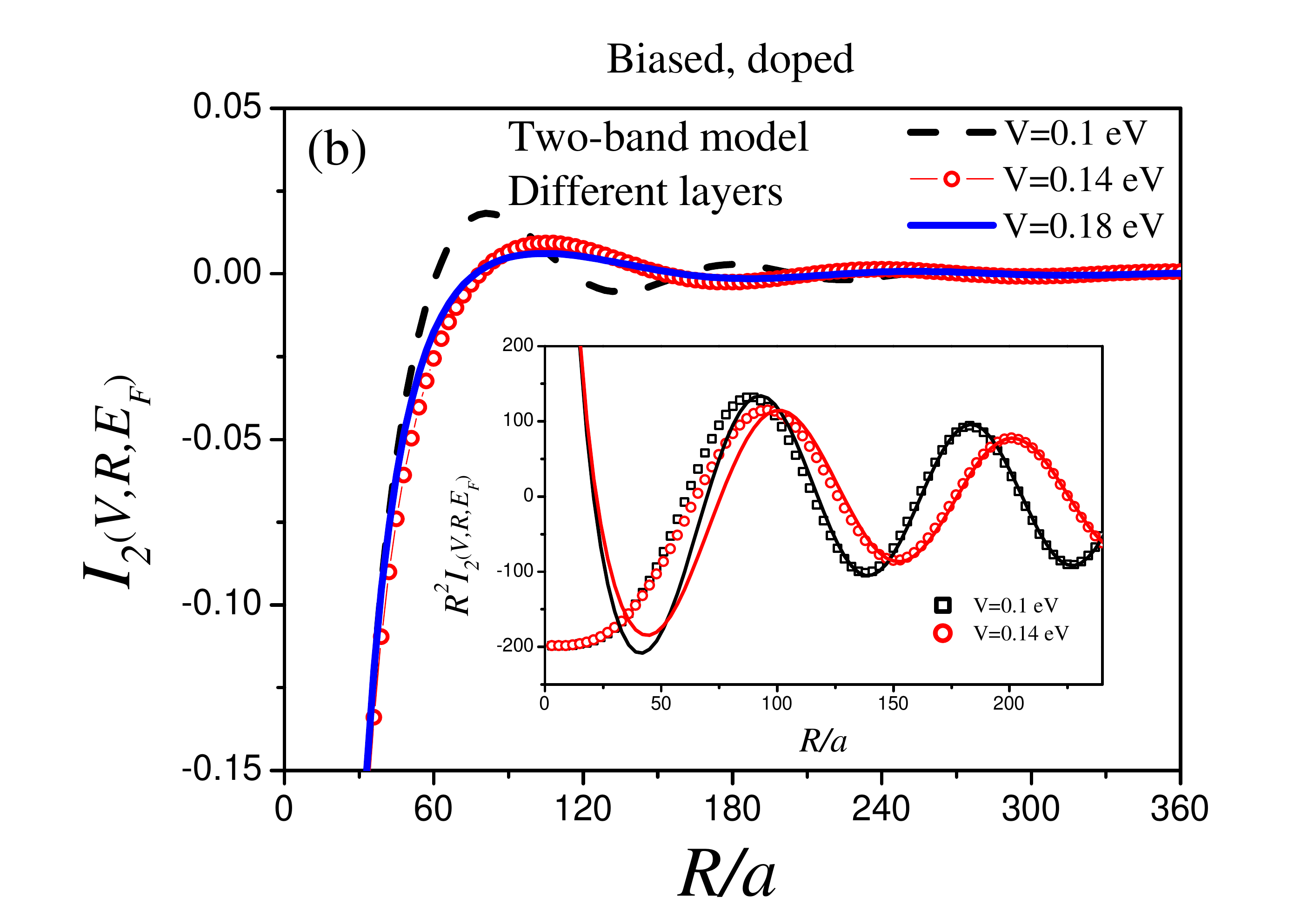}
\caption{ (Color online) (a) $I_1(V, R, E_{\rm F})$ as a function
of $R$ for different gate voltage at $E_{\rm F}=0.1$ eV. To
emphasis the amplitude value of $I_1(V, R, E_{\rm F})$, $R^2
I_1(V, R, E_{\rm F})$ is shown for different gate voltages. (b)
The same as (a) for $I_2(V, R, E_{\rm F})$. In the inset of (b),
solid lines refer to the analytical results of
Eq.~(\ref{eq:i2-full}) and are compared to the numerical
evaluation of Eq.~(\ref{eq:i2vef}), plotted as symbols, show their
difference at short distance while reaching each other quite well
as $R$ increases. Here,  $\alpha=0.42$, $\beta=1.6$ and
$\alpha=0.33$, $\beta=1.45$ for $V=0.1$ and $0.14$ eV,
respectively. \label{fig:Graph06}}
\end{figure}

\subsection{Biased BLG, doped and undoped}

By turning on the gate voltage perpendicular to the system, a new type of dispersion relation of the band is emerged. At zero Fermi energy, due to a gap opening and consequently removing the available energy states for the mediating electrons, the response function between electrons decreases, and as a result, the RKKY interaction decreases much faster than $R^{-2}$.

Fig.~\ref{fig:Graph03}(a) shows the integral $I_1(V, R, E_{\rm F}=0)$ as a function of $R$ [Eq. \eqref{eq:I1B1B1}] for different gate voltages in the two-band continuum model. The function $I_1(V,R, E_{\rm F}=0)$ decreases by increasing the bias voltage and oscillates slightly around its zero value. Decaying structures and oscillations depend on the bias voltage. Similarly, Fig.~\ref{fig:Graph03}(b) shows the integral $I_2(V,R, E_{\rm F}=0)$ as a function of $R$ [Eq. \eqref{eq:i2vef}] for different gate voltages. The function decays as $R$ increases and remains negative.

One goal of the present work is to understand the RKKY interaction in a doped BLG system. For this purpose, we consider a finite gate voltage together with the finite Fermi energy to calculate the RKKY interaction. Fig.~\ref{fig:Graph06} shows the integrals $I_1(V, R, E_{\rm F})$ and $I_2(V, R, E_{\rm F})$ as a function of $R$ for different gate voltages at given $E_{\rm F}=0.1$ eV. Our results show that $I_1(V, R, E_{\rm F})$ is sensitive to $V$ and by growing it around $2 E_{\rm F}$ values, the amplitude and also the wavelength of the oscillation of $I_1(V, R, E_{\rm F})$ increases. However, $I_2(V, R, E_{\rm F})$ slightly changes with the gate voltage. Similar to the case of unbiased and doped BLG, the integral $I_1$ for non-zero $V$ exhibits an oscillatory behavior as a function of $R$ with a period now controlled by both the Fermi momentum and the gate voltage. Our numerical results show that the period of the oscillations can be fitted quite well by $\pi/k_{\rm
V}$ where $k_{\rm V}=((2m E_{\rm F})^2-m^2V^2)^{1/4}$. One interesting feature in this case is that the long-range behavior of the RKKY interaction for the impurities on the same layer is similar that of a standard 2D electron gas. Another interesting feature is the enhancement of the RKKY interaction by increasing the gate voltage illustrated by ($R^2 I_1$) results in the inset of Fig.~\ref{fig:Graph06}(a). Since the RKKY interaction decays rapidly, it is almost difficult to measure it experimentally. Based on our results, here we proposes that the tuning of the gate voltage to a certain value, $2 E_{\rm F}$, will noticeably enhances the strength of the RKKY interaction and thereby makes it accessible for experimental probes.

Finally, we find that our numerical results for large distances between two impurities located on different layers [Eq. \eqref{eq:i2vef}] can be faithfully fitted by an analytical expression very similar to the unbiased case given in Eq. \eqref{eq:I2B1B2V0Asy}. This asymptotic fit is given by
\begin{widetext}
\begin{eqnarray}
\lim _{R \gg a} I_2(V,R,k_F)=-\frac{\pi}{2mR^2}\left[\sqrt{2}e^{-k_V R}\cos{(k_VR)}-\alpha \sin(2k_VR)-\beta\frac{\cos{(2k_VR)}}{k_VR}\right].
\label{eq:i2-full}
\end{eqnarray}
\end{widetext}
where $k_V=\left((2m E_{\rm F})^2-m^2V^2\right)^{1/4}$, $\beta$ and $\alpha$ are parameters controlled by $E_{\rm F}$ and $V$. In the inset of Fig.~\ref{fig:Graph06}(b), solid lines refer to the analytical results of Eq.~(\ref{eq:i2-full}) compared to the numerical evaluation of Eq.~(\ref{eq:i2vef}), plotted as symbols, showing their difference at short distance while reaching each other quite well as $R$ increases.

\begin{table*}[t!]
    \begin{center}
\begin{tabular}{l|c|c|r}
\hline\hline
\vline ~ References & System & RKKY interaction for same sublattice& RKKY interaction for different sublattices\vline \\
\hline
\vline ~ Ref.~[\onlinecite{RKKY-2D}]& 2DEG &  $R^{-2}\sin(2k_{\rm F}R)$ & - \vline\\
\hline
\vline ~ Ref.~[\onlinecite{Larsen}]& 2DEG+ impurity &  $R^{-2}\sin(2k_{\rm F}R)e^{-\alpha R}$  & - \vline\\
\hline
\vline ~ Ref.~[\onlinecite{MSashi1}] & SLG ($E_{\rm F} =0$) & $-R^{-3}\Phi_{AA}$ & $3R^{-3}\Phi_{AB}$ \vline\\
\hline
\vline ~ Ref.~[\onlinecite{MSashi2}] & SLG($E_{\rm F}\neq 0$) & $-R^{-2}\sin(2k_{\rm F}R) \Phi_{AA}$ & $R^{-2}\sin(2k_{\rm F}R) \Phi_{AB}$ \vline\\
\hline
\vline ~ Ref.~[\onlinecite{Kogan},\onlinecite{Jiang}] & BLG($E_{\rm F}= 0$, $V=0$) & $-R^{-2} \Phi_{AA}$ &  $R^{-2} \Phi_{AB}$ \vline\\
\hline
\vline ~ Present work & BLG($E_{\rm F}\neq 0$, $V=0$) & $-R^{-2}\cos(k_{\rm F}R)[e^{-k_{\rm F}R}+ 2^{-1/2}\sin(k_{\rm F}R)] \Phi_{B_1B_1}$ &$R^{-2}\cos(k_{\rm F}R)[e^{-k_{\rm F}R}- 2^{-1/2}\sin(k_{\rm F}R)]\Phi_{B_1B_2}$ \vline\\
\hline
\vline ~ Present work & BLG($E_{\rm F}\neq 0$, $V \neq 0$) & $-R^{-2}\sin(2l) \Phi_{B_1B_1}$ &$R^{-2}[e^{-l}\cos{(l)}-\alpha \sin(2l)-\beta\frac{\cos{(2l)}}{l}] \Phi_{B_1B_2}$ \vline\\
\hline\hline
\end{tabular}
\caption{A breakdown of the results on the scaling form of the RKKY interactions in two dimensional electron gas ( 2DEG), SLG and BLG. The RKKY interactions are proportional to values given in the third and fourth columns. $\alpha$ and $\beta$ are parameters controlled by $E_{\rm F}$ and $V$. The parameter $l=k_V R$ where $k_V=\left((2m E_{\rm F})^2-m^2V^2\right)^{1/4}$. The functions $\Phi_{AA}$ and $\Phi_{AB}$ are given by $ 1+\cos[(\bm K-\bm K')\cdot \bm R]$ and $1+\cos[(\bm K-\bm K')\cdot\bm R+\pi-2\theta_R]$, respectively.}
\label{table0}
\end{center}
\end{table*}

\section{Summary}

We have studied the effect of the bias voltage on the RKKY interaction in doped and gapped BLG. Our approach is based on the lattice Green's function technique. Near the Dirac points, charge carriers in BLG have parabolic energy spectrum with a finite density of states at zero energy, similar to the conventional non-relativistic electrons. On the other hand, these quasiparticles are also chiral and described by spinor wave functions. Therefore, the dependence of the RKKY interaction on the position vector $\bm R$ between two local magnetic moments is not only controlled by the dispersion relation but also by the chirality, which makes it directional-dependent as also shown to be the case for SLG ~\cite{MSashi1,MSashi2} by the phase factors $\Phi_{\alpha \beta}$.

Similar to SLG, we report the ferromagnetic interaction for moments on the same layers and anti-ferromagnetic coupling for those placed on the opposite layers in unbiased and undoped BLG. We associate this feature to the particle-hole symmetry and the bipartite nature of the lattice within the two-band model as argued in Ref.~[\onlinecite{Saremi}].

For the unbiased and doped case, we managed to find the analytical expressions of the RKKY interaction in terms of the Meijer G-functions and their long-range behavior was also reported. The salient feature of the asymptotic behavior is that the power-law decay $R^{-2}$ is accompanied by an exponential factor as $J_{B_1B_{1(2)}}\propto \mp R^{-2}\cos(k_{\rm F}R)[e^{-k_{\rm F}R}\pm 2^{-1/2}\sin(k_{\rm F}R)]$. It was shown that the mediating carriers of a gapped graphene \cite{Dugaev, KoganGapped} or SLG with disorder \cite{LeeSLGDisorder}  produce an exponential decay in the RKKY interaction; however, for a pristine unbiased BLG, which is gapless, we associate this exponential decay to the chiral nature of the carriers in the system.

We have supplemented the results from the two-band model for the unbiased case with our calculations using the four-band model to identify the validity of the two-band model and the discrepancy between both models. Within the four-band model, when the system is highly doped, the application of the two-band model is questionable. In this regime, the main features of the RKKY interaction are only captured by the four-band model. In low-energy region, we have shown that the two models are different at the short-range of the distance between impurities located on different layers. In addition, we have observed that the oscillations of $\chi_{B_1B_1}$ exhibit a beating pattern with two characteristic periods associated to the two Fermi momenta.

For the biased and doped BLG, we have shown that the gate voltage and the Fermi energy can vary independently to determine the RKKY interaction. One of the fascinating features in this case is the possibility for the enhancement of the interaction by tuning the gate voltage and/or the Fermi energy, which opens an avenue to probe the interaction experimentally. We have obtained the asymptotic behavior of the RKKY interaction analytically for each case and the expressions are given in Table~\ref{table0}. In order to compare the RKKY interaction in BLG with an ordinary 2DEG we have reported the interaction in clean 2DEG~\cite{RKKY-2D} and in the presence of disorder where the exponential decays is introduced \cite{Larsen}.

\section{Acknowledgments}
MS thanks Nico Temme for useful discussions on the Meijer G-functions. The work at the University of Missouri was supported by the U. S. Department of Energy through Grant No.
DE-FG02-00ER45818.

\end{document}